\documentclass[submitting]{nst}
\usepackage[percent]{overpic} 
\usepackage{adjustbox}
\usepackage{epstopdf}
\usepackage{subfigure,dcolumn}
\usepackage{mhchem}
\usepackage{upgreek}
\usepackage{hyperref}
\usepackage{tikz}
\usetikzlibrary{calc}

\makeatletter
\newcommand{\outsidenote}[1]{%
  \tikz[remember picture,overlay]{%
    \if@firstcolumn
      \node[
        anchor=east,
        text width=3cm,
        align=left,
        font=\small
      ]
      at ([xshift=2cm,yshift=-\pagetotal]current page.north west)
      {#1};
    \else
      \node[
        anchor=west,
        text width=3cm,
        align=left,
        font=\small
      ]
      at ([xshift=-1cm,yshift=-\pagetotal]current page.north east)
      {#1};
    \fi
  }%
}
\makeatother

\begin{document}

\title{Multiplexed SiPM Readout of Plastic Scintillating Fiber Detector for Muon Tomography}\thanks{Supported by the National Natural Science Foundation of China under Grant No. 12205174.}

\author{Chenghan Lv}
\affiliation{Institute for Advanced Technology, Shandong University, Jinan 250100, China}
\affiliation{Particle Physics Research Center, Shandong Institute of Advanced Technology, Jinan 250100, China}

\author{Kun Hu} 
\affiliation{Institute of Frontier and interdisciplinary Science, Shandong University, Qingdao 266237, China}

\author{Huiling Li}
\email[Corresponding author, ]{huiling.li@iat.cn}
\affiliation{Particle Physics Research Center, Shandong Institute of Advanced Technology, Jinan 250100, China}

\author{Hui Liang}
\affiliation{Particle Physics Research Center, Shandong Institute of Advanced Technology, Jinan 250100, China}

\author{Cong Liu}
\affiliation{Particle Physics Research Center, Shandong Institute of Advanced Technology, Jinan 250100, China}

\author{Hongbo Wang}
\affiliation{Particle Physics Research Center, Shandong Institute of Advanced Technology, Jinan 250100, China}

\author{Zibing Wu}
\affiliation{Institute for Advanced Technology, Shandong University, Jinan 250100, China}
\affiliation{Particle Physics Research Center, Shandong Institute of Advanced Technology, Jinan 250100, China}

\author{Weiwei Xu}
\email[Corresponding author, ]{Weiwei.Xu@email.sdu.edu.cn}
\affiliation{Institute for Advanced Technology, Shandong University, Jinan 250100, China}
\affiliation{Particle Physics Research Center, Shandong Institute of Advanced Technology, Jinan 250100, China}

\begin{abstract}
Muon tomography is a non-destructive imaging technique that uses cosmic-ray muons to probe dense materials. Bar scintillator and scintillating fiber detectors equipped with one-dimensional SiPM arrays offer compact, high-resolution solutions, but large-area implementations require effective reduction of readout channels while preserving detector performance.
To address this challenge, we present a novel multiplexing scheme based on a diode-based symmetric charge division circuit combined with a position-encoding algorithm, enabling up to $N_{\textrm{SiPM}}^{\textrm{max}}=C^{2}_{N_{\textrm{ele}}}$ SiPM channels to be read out using only ${N_{\textrm{ele}}}$ electronic channels. Circuit simulations confirm the feasibility of the multiplexing design and guide the choice of appropriate diodes to preserve SiPM signal integrity. The approach was validated using a SciFi detector module comprising 21 SiPM channels multiplexed into 7 electronic channels.
Electronic tests show that this multiplexing circuit exhibits low crosstalk between electronic channels, and preserves linearity over a dynamic range from $\sim$10 to 122 photoelectrons.
Cosmic-ray measurements further show that the multiplexed SciFi detector achieves a detection efficiency above 95\% and a spatial resolution of about 0.65~mm, with only minor degradation compared to the direct (per SiPM channel) readout. These results verify that the proposed method provides a scalable and cost-effective readout solution for large-area muon tomography systems and is applicable to other scintillator-based detectors employing similar one-dimensional SiPM array readout.

\end{abstract}


\maketitle

\section{Introduction}
\label{sec:introduction}
Cosmic-ray muons are secondary particles produced by the interactions of primary cosmic rays with the Earth’s atmosphere. These muons constitute a naturally available, highly penetrating source of radiation, with a typical flux of about 1 $\rm{cm^{-2}min^{-1}}$ and an average energy of approximately 4 GeV at sea level~\cite{Navas:2024}. By measuring either the scattering angles or transmission rates of muons from multiple directions, muon tomography enables non-destructive imaging of the internal structures of large and dense objects~\cite{Borozdin2003,Tang2020:MuonSourceReview,Bonomi:2020,Tanaka:2023}. This rapidly emerging technique has been applied across diverse fields, including nuclear waste monitoring~\cite{Jonkmans:2013NuclearWasteMuon,Mahon:2018}, border security~\cite{TUMUTY:2019, Xiao:2018, Luo:2022,He:2022MuonANN, Barnes:2023}, archaeology~\cite{Morishima:2017ghw,Liu:2023XianWall}, and geological exploration~\cite{Han:2020TunnelMuonFlux,Cheng:2022, Liu:2024,MuGridV2:2025}, among others, highlighting its broad applicability.

A major challenge in muon tomography is the relatively low cosmic-ray flux, which makes data acquisition time-consuming and limits the achievable imaging resolution. To maximize muon statistics and improve reconstruction quality, detector systems typically require large active areas and high spatial granularity. Plastic scintillator detectors are widely used for this purpose because they can be produced cost-effectively over large areas, offer high detection efficiency, and maintain stable performance under varying environmental conditions, making them well suited for large-scale muography applications.

Millimeter-level spatial resolution is commonly achieved by segmenting bulk scintillators into long bars with rectangular~\cite{Lesparre:2012,Dong:2018,MuGrid:2022} or triangular~\cite{Anstasio:2013, Hu:2020PlasticMuonDetector, Luo:2022CompactMuonImaging,Wang:2024} cross sections coupled to SiPMs. For submillimeter precision, plastic scintillating fiber (SciFi) detectors~\cite{Clarkson:2014,Anbarjafari:2021,Zhai:2022,Zhai:2024,Li:2025} are preferred due to their fine granularity and long attenuation length. These detectors are commonly read out using silicon photomultiplier (SiPM) arrays rather than PMTs to achieve compact module designs. However, increasing spatial resolution and detector area inevitably requires a large number of readout channels, leading to higher power consumption, system cost, and integration complexity. Due to the low flux of muon at the sea level, only a small portion of the detector channels is typically traversed by muons at any instant, this motivates the development of efficient channel-reduction (multiplexing) techniques
to balance performance and system scalability.

Existing multiplexing strategies can be categorized into two types: (i) optical multiplexing, in which multiple scintillating fibers are grouped before coupling to photodetectors, and (ii) electronic multiplexing, in which signals from photodetectors are combined before the front-end electronics. 

In the optical multiplexing strategy, multiple scintillating fibers are physically grouped before coupling them to MAPMT or SiPM channels~\cite{Clarkson:2014,Anbarjafari:2021} to reduce the number of electronic readout channels. Another optical multiplexing method that also relies on grouping fibers has been investigated, in which the grouping is implemented optically using additional masking layers~\cite{Sehgal2025OpticalMultiplexing}. However, both methods fundamentally depend on modifying the detector geometry to realize a specific multiplexing pattern. Consequently, any change in the multiplexing scheme necessitates physical reconfiguration of the detector, making these methods inherently inflexible for large-scale or frequently adjusted system designs.

In the electronic multiplexing strategy, the channel reduction is achieved entirely through the readout electronics, without altering the detector structure. This provides much greater flexibility when the multiplexing logic needs to be updated. Several electronic multiplexing circuits have been developed for TOF-PET systems~\cite{Park:2022}. The resistive charge-division networks~\cite{Siegel:1996}, which achieve channel reduction through analog charge sharing, are suited to crystal scintillator detectors with high light yield and moderate count rate, where the impact of RC delay and electronic noise are less critical. Time- ~\cite{Won:2016} or frequency-modulation~\cite{Wonders:2020} based multiplexing techniques encode channel identity using signal delays or spectral features, and typically rely on precise timing control and fast electronics, resulting in increased signal-processing complexity. Digital or highly digitized SiPM readout architectures are mainly developed to optimize timing performance, particularly for time-of-flight measurements. In general, these approaches are tailored to two-dimensional SiPM arrays coupled to crystal scintillators. Such design assumptions differ markedly from those of bar scintillator or scintillating-fiber detectors, which typically employ one-dimensional SiPM readout and operate under low-light conditions, with signal amplitudes ranging from a few to several hundred photoelectrons per event. As a result, existing TOF-PET multiplexing designs cannot be readily applied, highlighting the need for a multiplexing solution optimized for low-light, one-dimensional detector configurations.
%Several electronic multiplexing circuits have been developed for TOF-PET systems~\cite{Park:2022}, where the designs are tailored to two-dimensional SiPM arrays and the high light yield of crystal scintillators. These conditions differ markedly from those of bar scintillator or SciFi detectors, which typically employ one-dimensional SiPM readout and produce relatively low light output ranging from a few to several hundred photoelectrons per event. Therefore, existing TOF-PET multiplexing designs cannot be readily applied, highlighting the need for a multiplexing solution optimized for low-light, one-dimensional detector configurations.

In this work, we develop a multiplexing method specifically designed for bar scintillator and SciFi detectors employing one-dimensional SiPM readout. The proposed scheme is based on a diode-based symmetric charge-division circuit combined with a position-encoding algorithm. Using a SciFi detector as a representative case, circuit simulations are first performed to validate the multiplexing concept and guide diode selection. Electronic measurements are then conducted to characterize the crosstalk and linearity of the multiplexing circuitry. Finally, cosmic-ray tests are carried out to evaluate the detection efficiency and spatial resolution of the multiplexed SciFi detector and compare them with those of a direct-readout configuration, demonstrating the practicality and scalability of the proposed approach for large-area muon imaging systems.

\section{Plastic scintillating fiber detector}
\begin{figure}[!htb]
%\centering
\includegraphics[width=0.8\hsize]{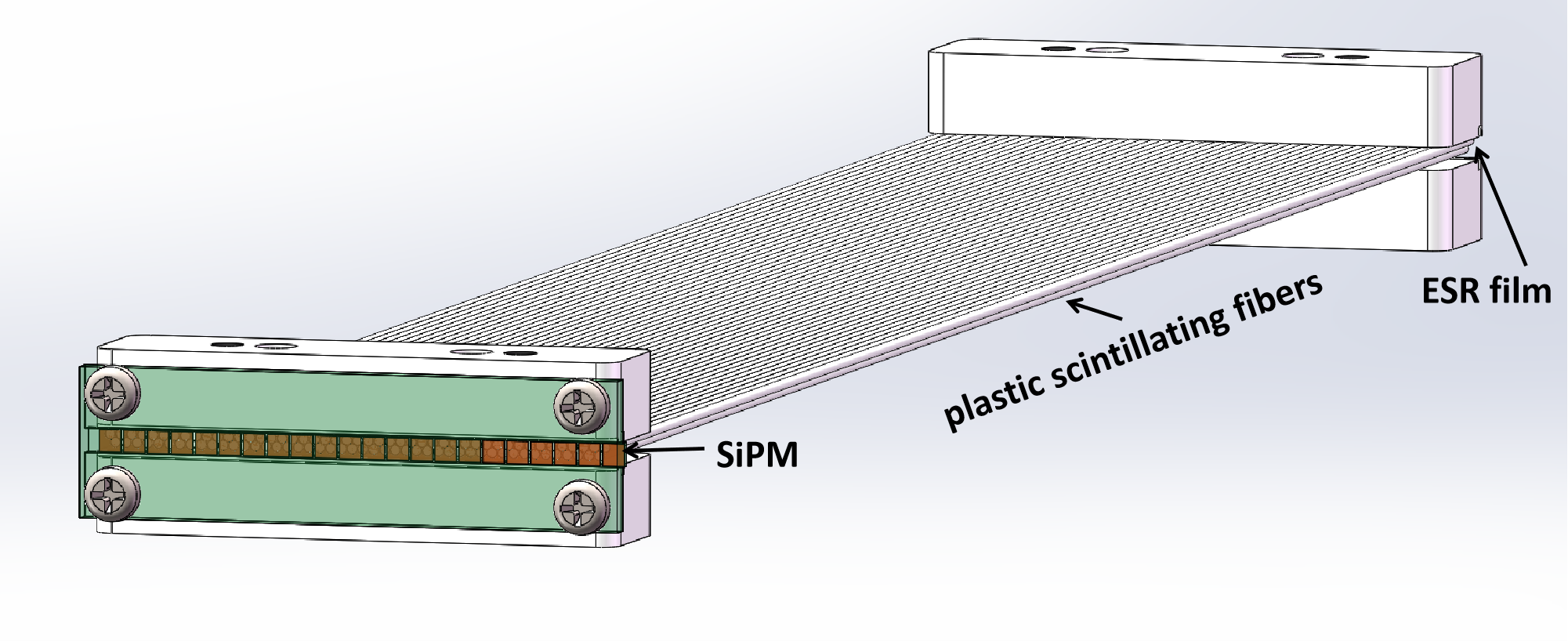}
\caption{Design of a SciFi module, consisting of two staggered layers of 1 mm diameter fibers read out by a one-dimensional SiPM array with 22 channels of $2\times2$~mm channel size.}
\label{fig:det}
\end{figure}

To demonstrate the proposed multiplexing strategy, we employ a SciFi detector module developed in our laboratory~\cite{Li:2025}, as illustrated in Fig.~\ref{fig:det}. The module is representative of typical bar- or fiber-based detectors that feature one-dimensional SiPM readout and relatively low light output, making it well suited for evaluating the performance and feasibility of the multiplexing scheme introduced in this work.

The detector provides submillimeter spatial resolution within a compact geometry suitable for muon tomography applications. It has a sensitive area of $\rm 5.5\times100~cm^{2}$ and consists of two staggered layers of double-clad round plastic scintillating fibers (Kuraray SCSF-78M), each 1~mm in diameter. One end of the fiber bundle is directly coupled to a one-dimensional SiPM array composed of 22 Hamamatsu S13360-2050VE devices with $2\times2$~mm channel size and 0.5~mm gap between adjacent SiPMs. The opposite end of the fibers is covered with an enhanced specular reflector (ESR) film to improve light collection and increase the overall signal yield of the module.

When a cosmic-ray muon traverses the SciFi module, scintillation light with wavelengths between 400 and 700~nm is generated and guided through the fibers onto the SiPM surface. Typically, one signal cluster corresponds to one or two adjacent SiPM channels fired within a single event. In some cases, optical crosstalk between neighboring fibers can result in three adjacent channels being fired simultaneously. The most probable value of the cluster signal is about 36 photoelectrons (p.e.) as measured in the muon test. To suppress background events, muon hits are selected by requiring the cluster signal to exceed 8~p.e.. The hit position is then reconstructed using a charge-weighted center-of-gravity method based on the fired channels:
\begin{equation}
x_{\mathrm{cog}} = \frac{\sum_{i} N_{i} x_{i}}{\sum_{i} N_{i}} ,
\end{equation}
where $x_{i}$ is the center position of the $i$th SiPM in the cluster, and $N_{i}$ is its corresponding signal size. This center-of-gravity approach is commonly employed in bar scintillator and SciFi detectors for position reconstruction.

\begin{figure}[!htbp]
\center
\includegraphics[width=0.9\hsize]{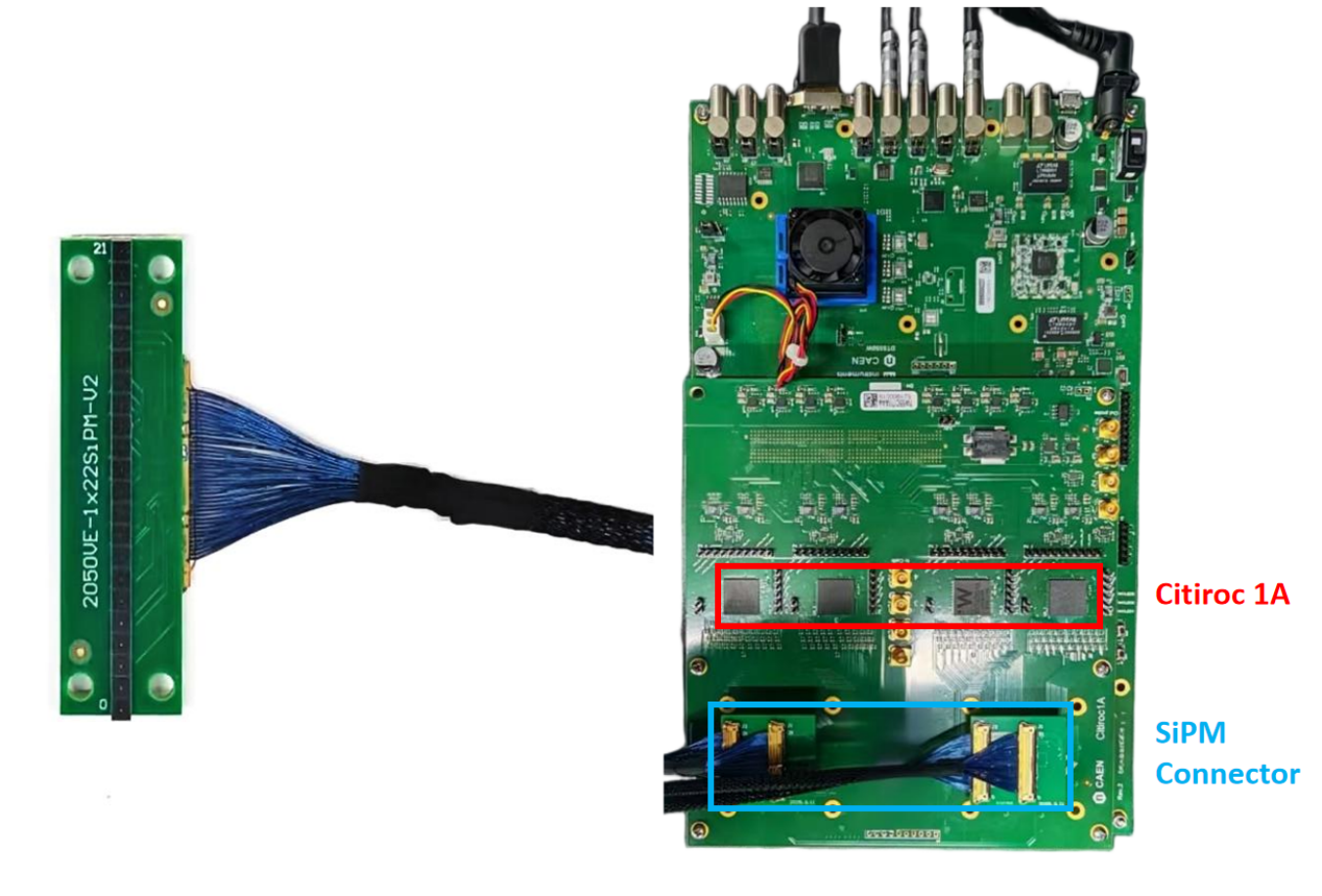}
\caption{ (Left) A 22-channel SiPM array and (right) its readout electronics DT5550W based on Citiroc 1A ASICs.}
\label{fig:sipmDAQ}
\end{figure}

For charge integration and digitization of SiPM signals, front-end electronics (DT5550W, CAEN) based on the Citiroc~1A ASIC~\cite{Citiroc1A} are employed. As shown in Fig.~\ref{fig:sipmDAQ}, the DT5550W board interfaces with up to four SiPM arrays and provides bias voltage through micro-coaxial cables. Each SiPM array is connected to one Citiroc~1A chip, where the analog pulses are amplified and integrated using a slow shaper and a peak-sensing technique. For each input channel, the Citiroc 1A provides both low-gain (LG) and high-gain (HG) analog outputs, covering a dynamic range of the SiPM signal up to 2500 p.e. The outputs are subsequently digitized by a 14-bit ADC and sent to an FPGA for packaging. Finally, the digitized SiPM charges are read out by the data acquisition computer.

\section{Multiplexing electronics}
To reduce the number of readout channels while maintaining detection efficiency and spatial resolution of the detector in muon tomography applications, a multiplexing scheme based on a diode-based symmetric charge-division circuit and a position-encoding algorithm has been developed. The corresponding multiplexing board is inserted between the SiPM arrays and the front-end electronics, allowing pulses from multiple SiPM channels to be combined and encoded before being processed. To demonstrate the principle and performance of the multiplexing electronics, a 22-channel SiPM array from one SciFi module is used in the following studies.

\subsection{Multiplexing method}
\begin{figure}[!htbp]
\center
\includegraphics[width=\hsize]{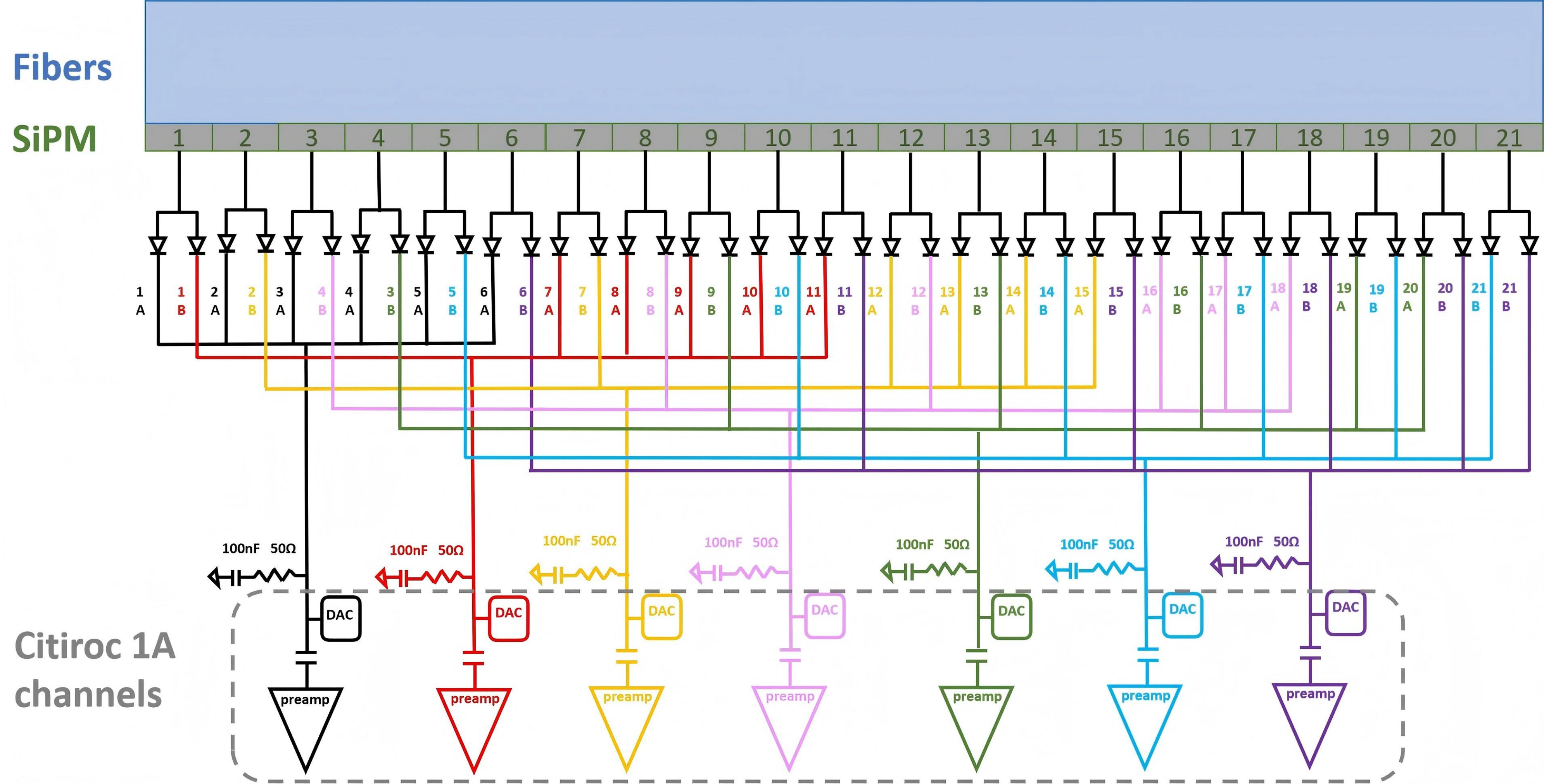}
\caption{Schematic of the multiplexed SiPM readout, where the signal of each SiPM channel is symmetrically divided into two parts by the diodes, and 21 SiPM channels are encoded into 7 electronic channels.}
\label{fig:scheme}
\end{figure}

Figure~\ref{fig:scheme} illustrates the concept of the multiplexed SiPM readout. In this method, each SiPM output signal is split into two components by a diode-based Symmetric Charge-Division (SCD) circuit~\cite{Jung:2021,Massari:2016}. These two components are then combined with signals from other SiPM channels to form two separate electronic readout channels. Compared with the conventional resistor-based SCD scheme~\cite{Olcott:2005,Wang:2016_SiPMmultiplex}, the inherent unidirectionality of the diodes effectively suppresses crosstalk between electronic channels. 

In designing the multiplexing circuit, special attention is paid to the requirements imposed by the center-of-gravity position reconstruction. Accurate hit-position determination relies on the relative amplitudes of signals from adjacent SiPM channels. In a multiplexed readout scheme, the SiPM signals are first combined through a passive multiplexing network before being decoded. During this process, SiPM signal may be affected by the inter-channel crosstalk or waveform distortion due to loading effects and the parasitic impedance of the multiplexing circuitry. To preserve the relative charge information of the original SiPM signals, the diode-based SCD scheme is chosen because of its inherent suppression of inter-channel crosstalk and its ability to minimize waveform distortion. This preservation of signal characteristics ensures compatibility with the subsequent charge-integration front-end electronics.
% , enabling more reliable calculation of the hit position using the standard center-of-gravity method.

The number of electronic channels, denoted as $N\mathrm{_{ele}}$, required for reading out $N\mathrm{_{det}}$ SiPM channels can be reduced by applying a position-encoding method. The mapping between each SiPM channel and the electronic channels is constructed according to the following criteria:
\begin{itemize}
\item Each SiPM channel is connected to two electronic channels, whereas each electronic channel is shared among multiple SiPM channels.
\item Any pair of electronic channels corresponds uniquely to one SiPM channel index.
\item All SiPM channels are covered by the established mapping relations.
\end{itemize}

Therefore, the maximum number of SiPM channels that can be handled by $N_\mathrm{{ele}}$ electronic channels is given by the number of combinations of any two electronics channels:
\begin{equation}
N_{\rm max} = C_{N\rm{_{ele}}}^{2} .
\end{equation}

\begin{figure}[!htbp]
\centering
\includegraphics[width=\hsize]{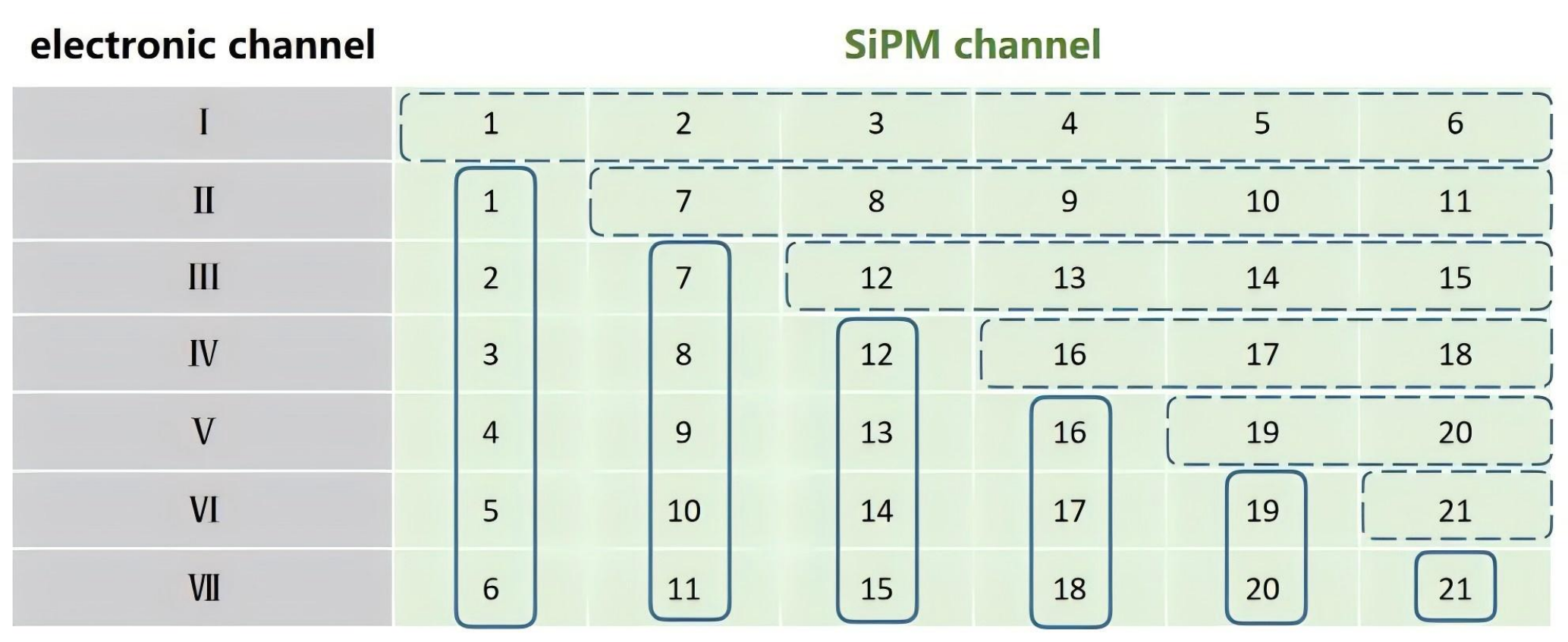}
\caption{Encoding map of 21 SiPM channels and 7 electronic readout channels. }
\label{fig:encoding}
\end{figure}

Since a signal cluster generated in a single event typically involves one or several adjacent SiPM channels, an encoding mapping table can be systematically constructed, as illustrated in Fig.~\ref{fig:encoding}. In this scheme, the first $N_\mathrm{ele}-1$ SiPM channels (indices 1–6) are grouped and connected to the first electronic channel. The first SiPM channel (index 1) is then used as a common reference between the first and second electronic channels. Another $N_\mathrm{ele}-2$ SiPM channels (indices 7–11) are selected from the remaining channels and, together with the common reference channel, are connected to the second electronic channel.

Similarly, the second SiPM channel of the first electronic channel (index 2) and the second SiPM channel of the second electronic channel (index 7) are used as common references for constructing the third electronic channel, to which an additional $N_\mathrm{ele}-3$ SiPM channels (indices 12–15) are connected. This iterative procedure continues until all SiPM channels are assigned, ensuring that each electronic channel shares at least one common SiPM channel with its neighbors to allow unique decoding.

If the total number of SiPM channels exceeds $N\mathrm{_{max}}$ for a given $N_{\mathrm{ele}}$, the additional SiPM channels can either be connected individually to available electronic channels, or the mapping can be extended by introducing one additional electronic channel together with a set of dummy SiPM inputs to maintain the encoding pattern. For the 22-channel SiPM array used in this work, we adopted a configuration of seven multiplexed electronic channels plus one individual channel. 

In practice, the achievable multiplexing factor is constrained by the noise performance of the combined readout. Because multiplexing merges the outputs of multiple SiPM channels, the baseline noise inevitably increases after encoding. Thus, the maximum practical multiplexing rate must be determined by examining the separation between the multiplexed noise level and the expected detector signal amplitude. In other words, the multiplexing factor should be chosen such that the elevated noise fluctuations do not obscure or distort the physical signals of interest.

By neglecting the crosstalk between electronic channels in the diode-based SCD circuits, the amplitude of a multiplexed electronic channel can be approximated as half of the sum of the corresponding SiPM signals defined in the mapping relation. After pedestal subtraction from the electronic channels, the first step of the decoding process is to identify the seed SiPM channel index $i$ by selecting the two electronic channels with the largest signal amplitudes. Using this seed channel index, its neighboring SiPM channels ($i-1$, $i+1$) are examined to determine whether they are also fired by checking their corresponding electronic channels. The number of adjacent channels to be searched depends on the expected cluster size of the detector. If any electronic channels still contain signals that cannot be associated with the identified cluster, the decoding procedure is repeated until all signals are assigned.

Finally, the signal values of all fired SiPM channels are obtained by solving the set of linear equations derived from the observed signals:
\begin{equation}
e_{k} = \frac{1}{2}\sum_{j}d_{j}
\end{equation} 
where $e_{k}$ is the signal amplitude of the $k$-th electronic channel, and $d_{j}$ denotes the signal of the $j$-th fired SiPM channel. 

This decoding strategy can reliably reconstruct events containing a single cluster, while events with multiple clusters are difficult to resolve due to channel ambiguity. Therefore, the proposed multiplexing scheme is suitable for detectors operating under low event-rate conditions. 
Although demonstrated with the SciFi detector, the same multiplexing principle can be applied to other scintillator-based detectors that employ comparable one-dimensional SiPM array readout configurations.

\subsection{Circuit simulation}
\begin{figure}[!htbp]
\center
\includegraphics[width=0.8\hsize]{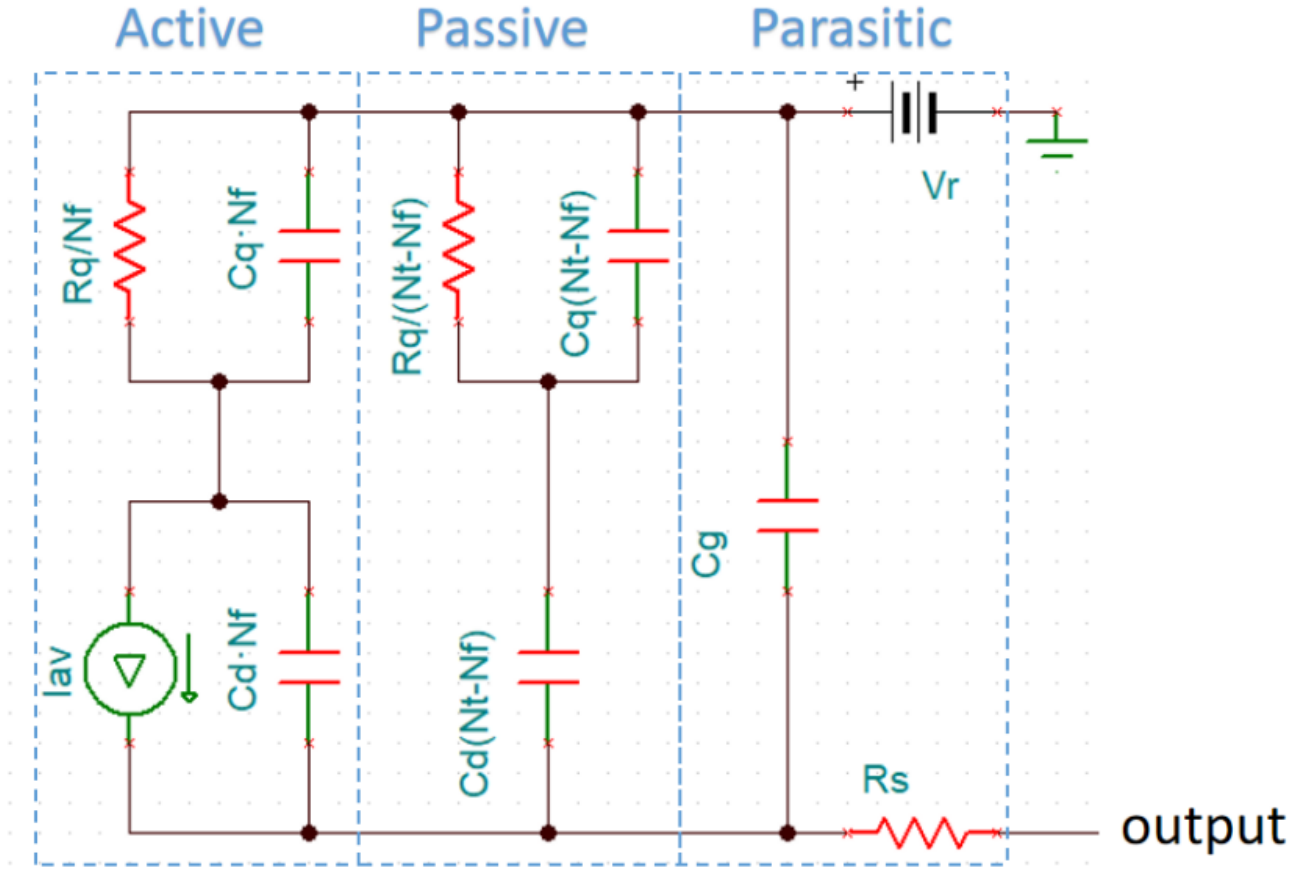}
\caption{The equivalent circuit of one SiPM channel. }
\label{fig:sipm}
\end{figure}

To evaluate the feasibility of the proposed multiplexing circuit and to select an appropriate diode, circuit simulations were performed using the TINA-TI tool~\cite{TINA-TI}. The equivalent circuit model of a single SiPM channel~\cite{Acerbi:2019SiPMs} was constructed based on the specifications of the Hamamatsu S13360-2050VE device provided by the manufacturer. As shown in Fig.~\ref{fig:sipm}, the SiPM model consists of three parts: an active component representing the triggered microcells, a passive component corresponding to the remaining non-triggered cells, and parasitic elements primarily arising from the metal grid and bonding pads. 
By varying the number of triggered microcells ($N_{\mathrm{f}}$), the model can reproduce SiPM signals over a wide range of incident light intensities.

\begin{figure}[!htbp]
\center
\includegraphics[width=0.9\hsize]{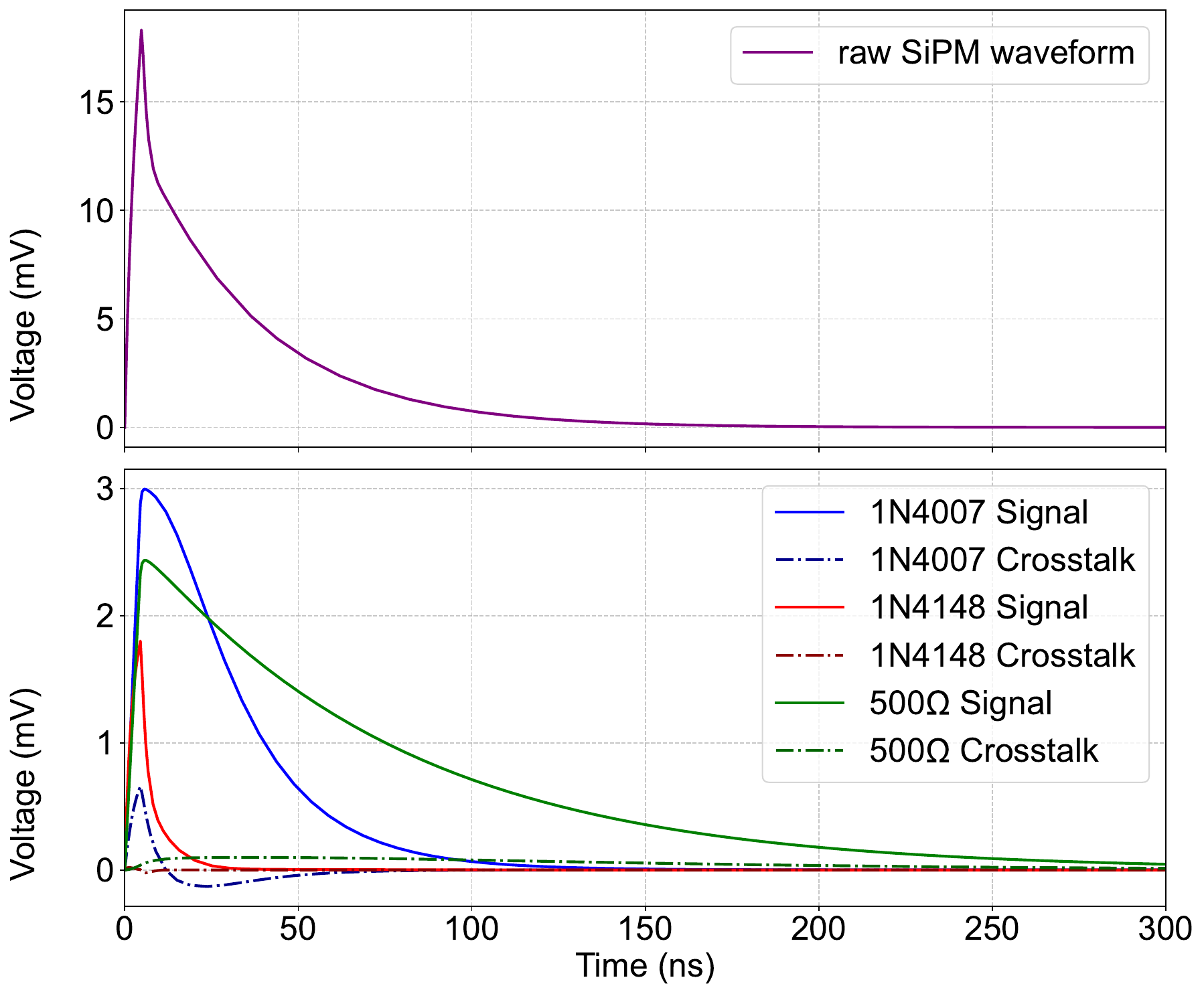}
\caption{ (Top) Simulated SiPM waveform corresponding to 20 fired microcells in one channel. (Bottom) Multiplexed output waveforms of signal and crosstalk channels for three circuit configurations.}
\label{fig:simPulse}
\end{figure}

The multiplexing circuit of 21 SiPM channels was simulated under three configurations: (1) a diode-based SCD circuit using high-speed switching diodes (1N4148), (2) a diode-based SCD circuit using silicon rectifier diodes (1N4007), and (3) a conventional resistor-based SCD circuit employing 500~$\Omega$ resistors. Based on the cluster signals of the SciFi detector, the simulation was performed with one SiPM channel generating signals corresponding to 20 fired microcells, as shown in the upper panel of Fig.~\ref{fig:simPulse}. The resulting multiplexed output waveforms are presented in the lower panel. The two multiplexed outputs connected to the fired SiPM channel were defined as signal channels, whereas the remaining outputs were used to evaluate the crosstalk behaviors.

The resistor-based SCD circuit exhibited significantly broader signal and crosstalk pulses than both diode-based configurations, deviating considerably from the raw SiPM waveform. Between the two diode options, the 1N4148 circuit produced narrower pulses with faster rise times but noticeably reduced amplitudes due to its smaller junction capacitance. In contrast, the 1N4007-based circuit generated multiplexed signals with larger amplitudes and a waveform shape that more faithfully preserves the charge-related characteristics of the original SiPM pulse, which is critical for detectors where integrated charge is the primary observable. Moreover, the larger signal amplitude of the 1N4007 configuration helps maintain a higher signal-to-noise ratio, while its higher junction capacitance contributes less to noise amplification when multiple channels are combined. Although both diode-based circuits exhibited bipolar crosstalk features that can be further suppressed by shaping amplification, the better waveform preservation, higher signal amplitude, and more favorable noise performance of the 1N4007 configuration make it the preferred choice for the multiplexing circuit in this work.

\section{Electronics performance}
\begin{figure}[!htbp]
 \centering
    % --------- 图 (a) ---------
    \begin{overpic}[width=\hsize]{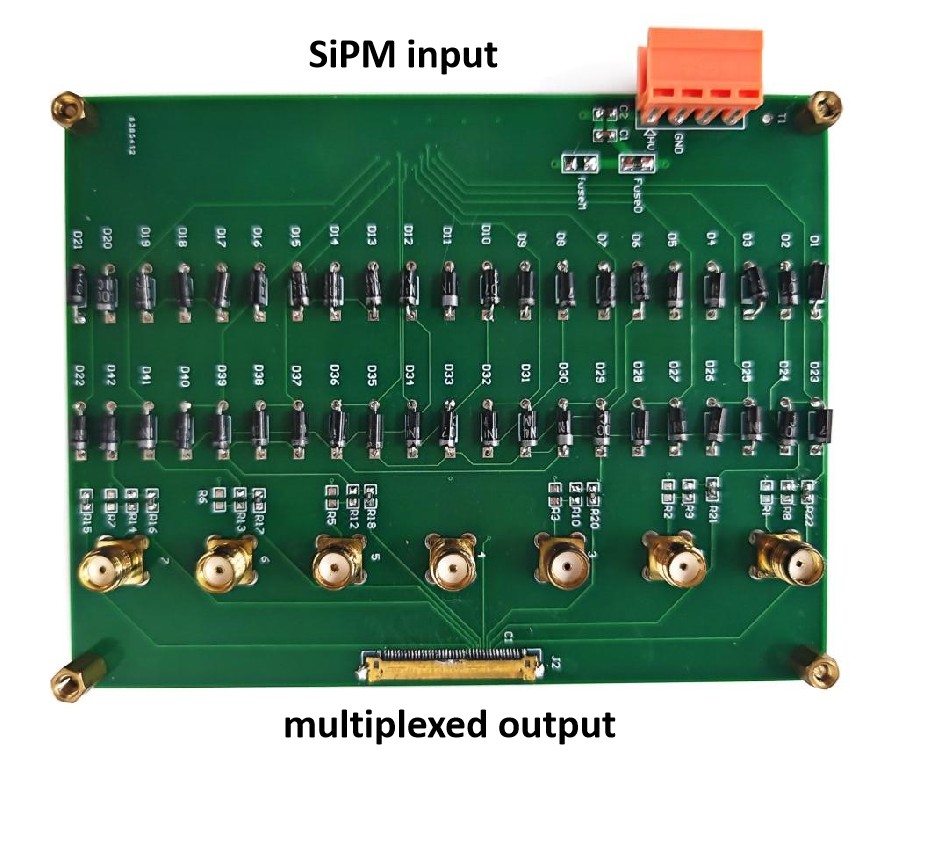}
        \put(1,80){\large\textbf{(a)}}  
    \end{overpic}

    \vspace{0.8em}

    % --------- 图 (b) ---------
    \begin{overpic}[width=\hsize]{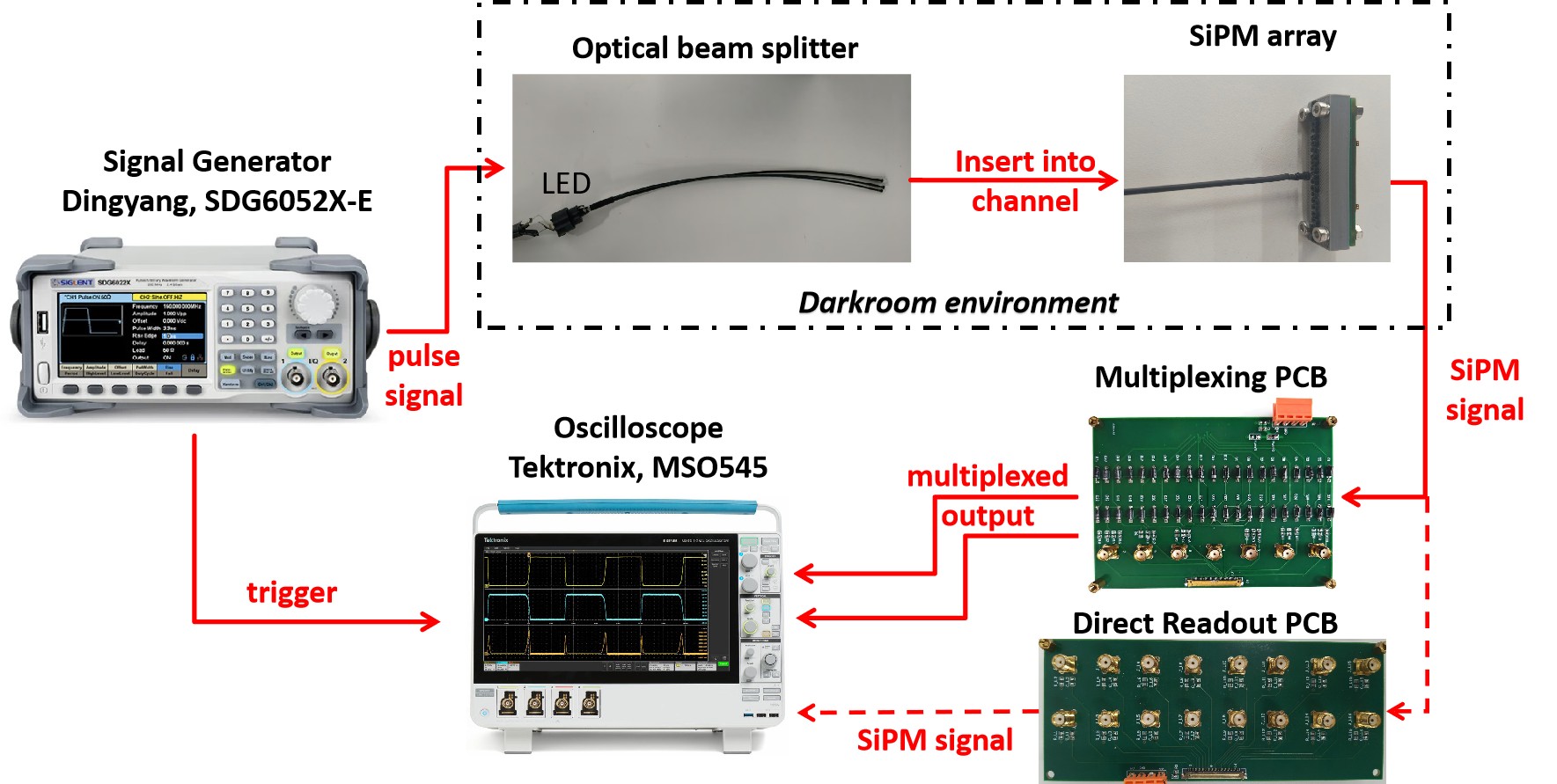}
        \put(1,60){\large\textbf{(b)}}  
    \end{overpic}
    \caption{(a) The multiplexing board to encode 22 SiPM channels into 7 multiplexed outputs and 1 direct output.  (b) Experimental setup to test the multiplexing circuit performance.}
    \label{fig:eleSetup}
\end{figure}

Following the circuit simulations and the selection of the 1N4007 diode as the preferred device, the performance of the implemented multiplexing board was experimentally evaluated in the setup as shown in Figure~\ref{fig:eleSetup}. A dedicated beam splitter, consisting of a UV LED and three scintillating fibers, was employed to generate controlled optical signals. A 22-channel SiPM array was equipped with a customized light shield that only allows the fibers from the beam splitter to illuminate specific channels, thereby simulating event clusters. Both the multiplexed output waveforms and the raw SiPM waveforms were monitored using an oscilloscope with the input trigger signal synchronized to the LED driving pulse.

After verifying the correct operation of the multiplexing logic on the board, tests were conducted to validate that the multiplexing circuit preserves the signal integrity required for reliable decoding and does not introduce significant crosstalk, non-linearity, or waveform distortion beyond the expectations of the circuit model.

\subsection{Signal and crosstalk}
\begin{figure}[!htbp]
 \centering
    % --------- 图 (a) ---------
    \begin{overpic}[width=\hsize]{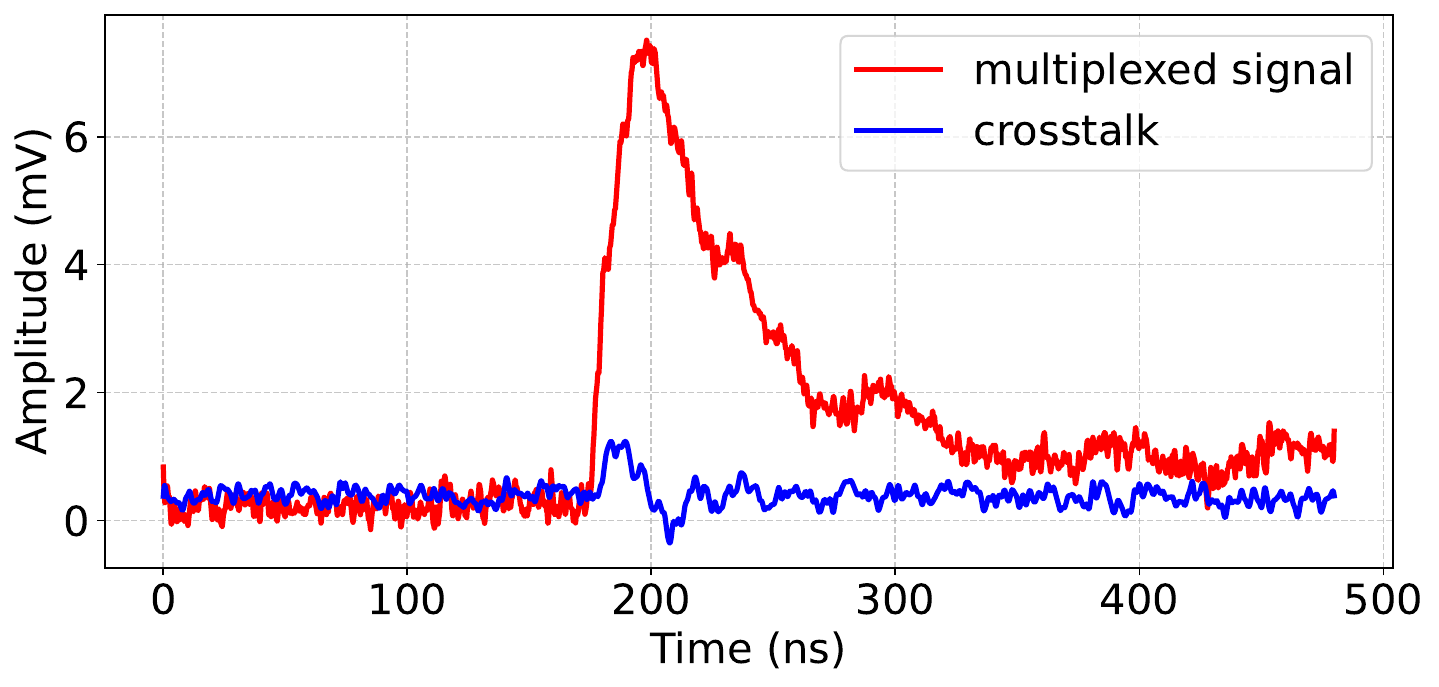}
        \put(10,42){\large\textbf{(a)}}  
    \end{overpic}

    \vspace{0.8em}

    % --------- 图 (b) ---------
    \begin{overpic}[width=\hsize]{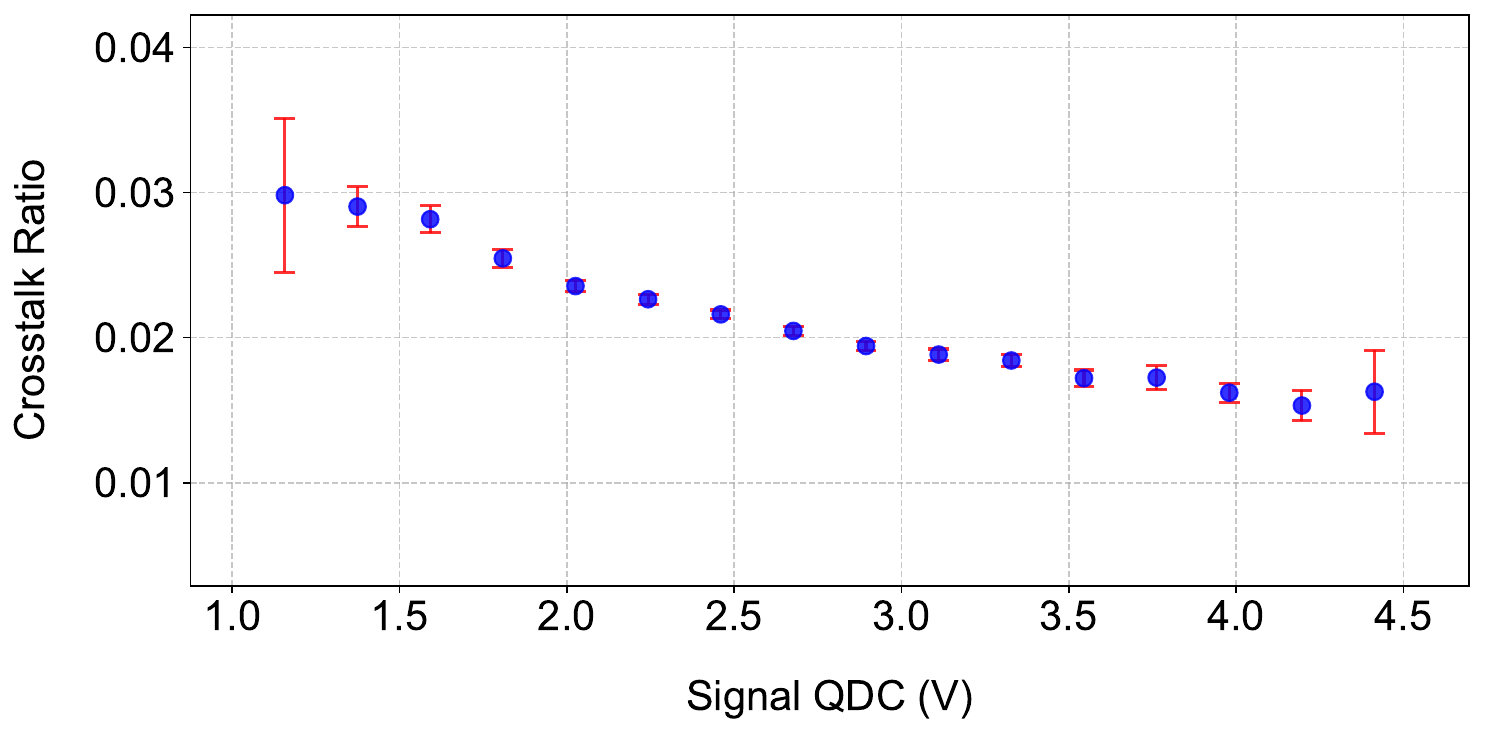}
        \put(15,44){\large\textbf{(b)}}  
    \end{overpic}
    \caption{(a) An example of output multiplexed signal and crosstalk waveforms, where the signal QDC value is about 2.7~V.  (b) Crosstalk ratio as a function of multiplexed signal charge.}
    \label{fig:xtratio}
\end{figure}

By selectively illuminating individual SiPM channels using the beam splitter, the multiplexed output and crosstalk waveforms of the 1N4007-based circuit were measured, as shown in Fig.~\ref{fig:xtratio}(a). The observed pulse shapes for both the signal and crosstalk outputs are consistent with the simulation results. The signal waveform exhibits a width comparable to that of the raw SiPM waveform, whose peak amplitude is approximately 60~mV on average.

The crosstalk ratio is defined as the ratio of the sum amplitude of the positive component of the crosstalk waveform to that of the multiplexed signal waveform within a 140~ns integration window. Figure~\ref{fig:xtratio}(b) shows the crosstalk ratio as a function of the multiplexed signal charge (QDC). The diode-based multiplexing circuit achieves a low crosstalk ratio below 3\% across the tested signal range.

\subsection{Linear response of signals}
Since the multiplexed signals are later used in the charge-weighted position reconstruction, it is essential that the multiplexing circuit maintains a linear response to the raw SiPM waveforms over the full dynamic range of the SciFi detector, from a few to several hundred photoelectrons.

To evaluate the linearity, the amplitude of the LED driving pulse was varied, and both the raw SiPM waveforms and the corresponding multiplexed output signals were recorded using an oscilloscope. For each amplitude setting, the number of photoelectrons ($N_{\mathrm{pe}}$) for both the raw and multiplexed signals was determined by fitting their QDC spectra with a modified Poisson distribution:
\begin{equation}
f(x) = K \cdot \frac{\lambda^{x/g} \cdot e^{-\lambda}}{\Gamma\left(\frac{x}{g} + 1\right)},
\label{eq:poisson}
\end{equation}
where $K$ is a coefficient, $x$ is the measured charge, $g$ is the SiPM gain, and $\lambda$ represents the mean number of detected photoelectrons $N_{\mathrm{pe}}$. 

\begin{figure}[!htbp]
    \centering

    % --------- 图 (a) ---------
    \begin{overpic}[width=\hsize]{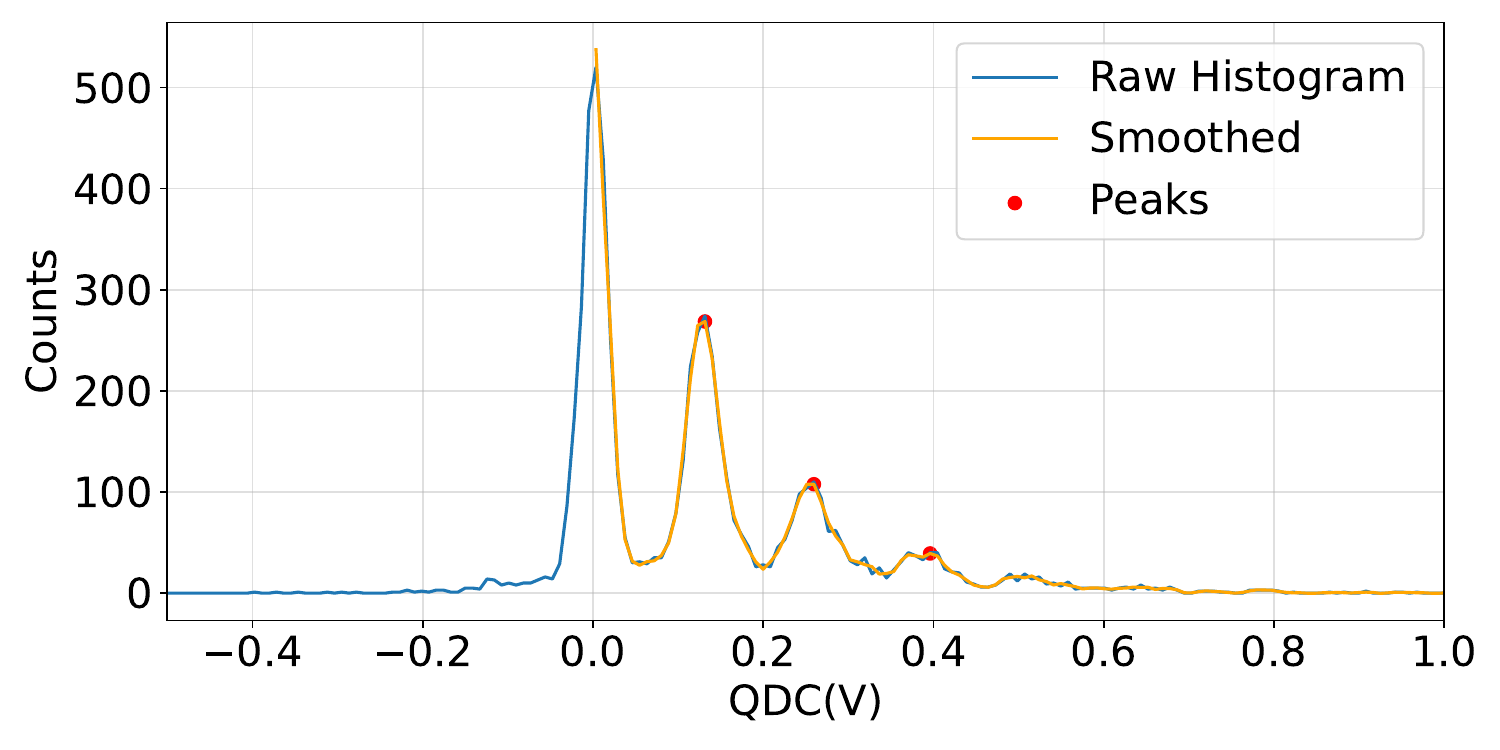}
        \put(15,40){\large\textbf{(a)}}  
    \end{overpic}

    \vspace{0.8em}

    % --------- 图 (b) ---------
    \begin{overpic}[width=\hsize]{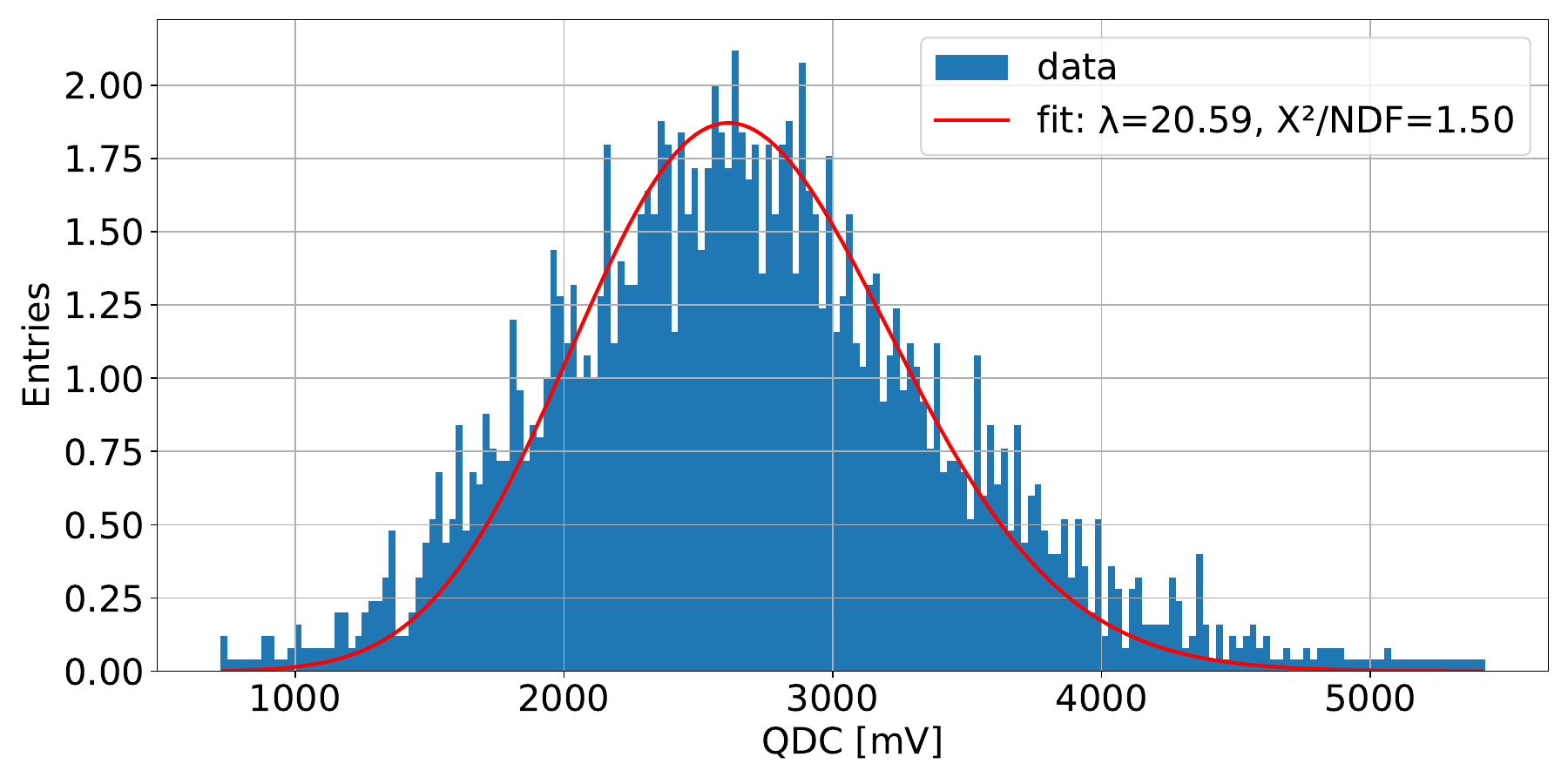}
        \put(15,40){\large\textbf{(b)}}  
    \end{overpic}

    \begin{overpic}[width=\hsize]{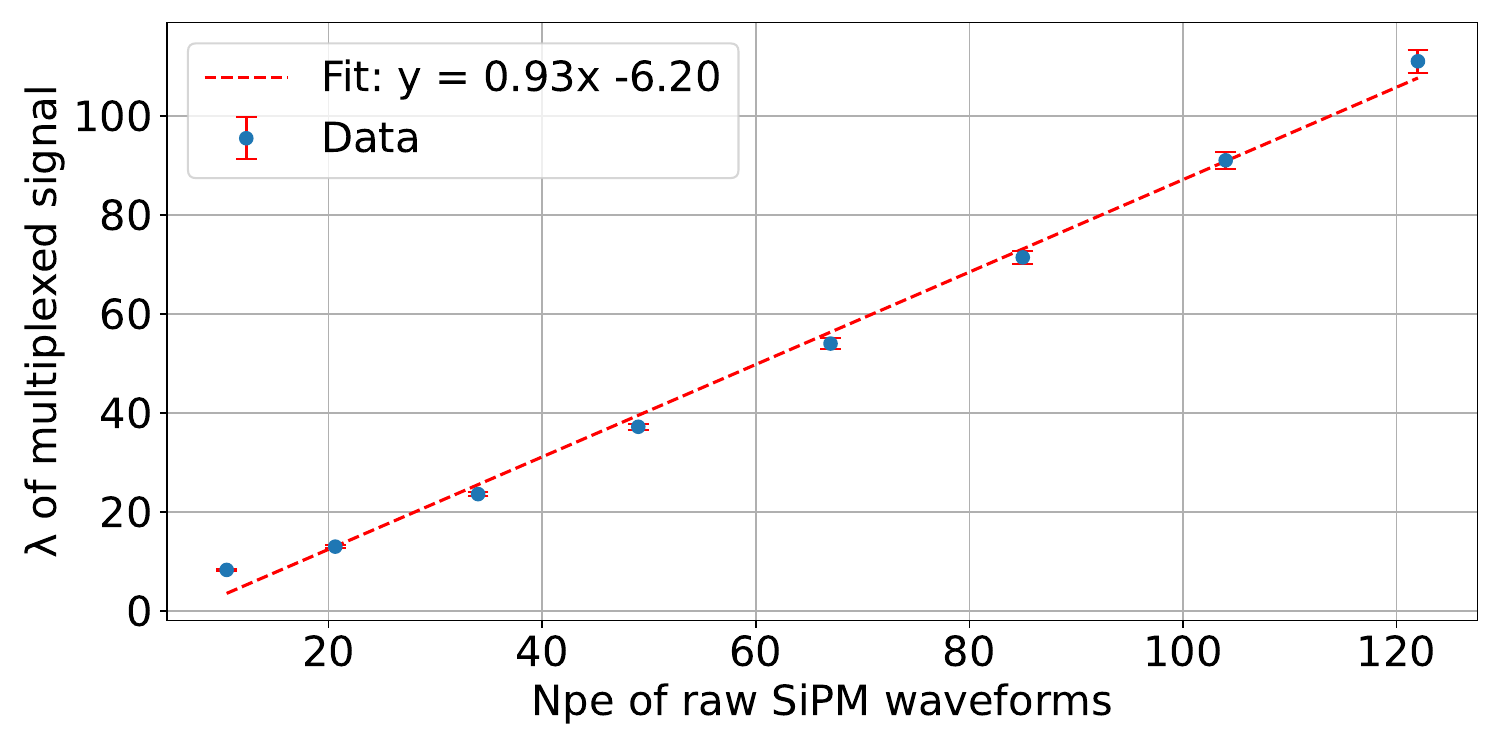}
        \put(15,30){\large\textbf{(c)}}  
    \end{overpic}
    
    \caption{(a) SiPM gain estimated from the charge spectrum. (b) Fit of $N_{\mathrm{pe}}$ of raw SiPM signals using the fixed gain value. (c) Linear relationship between the fitted $N_{\mathrm{pe}}$ of the multiplexed signals and the raw SiPM signals.}

    \label{fig:npe}
\end{figure}

The SiPM gain $g$ was determined in advance from the charge spectrum of raw waveforms under low-light conditions and fixed during the fitting of the raw SiPM spectra. As shown in Fig.~\ref{fig:npe}(a), $g$ is estimated from the mean QDC difference between adjacent single-photoelectron peaks. Using this fixed gain, $N_{\mathrm{pe}}$ of the raw SiPM signals was extracted for each LED amplitude setting, as illustrated in Fig.~\ref{fig:npe}(b).

The relationship between the fitted $N_{\mathrm{pe}}$ of the multiplexed signals and the raw SiPM signals is shown in Fig.~\ref{fig:npe}. The data are fitted with a straight line, demonstrating that the multiplexing circuit preserves an approximately linear response for SiPM signals from about 10~p.e. up to 122~p.e., which covers the majority operational signal range of the SciFi module used in this work.

\section{Detector performance}
\begin{figure}[!htbp]
\centerline{\includegraphics[width=0.9\hsize]{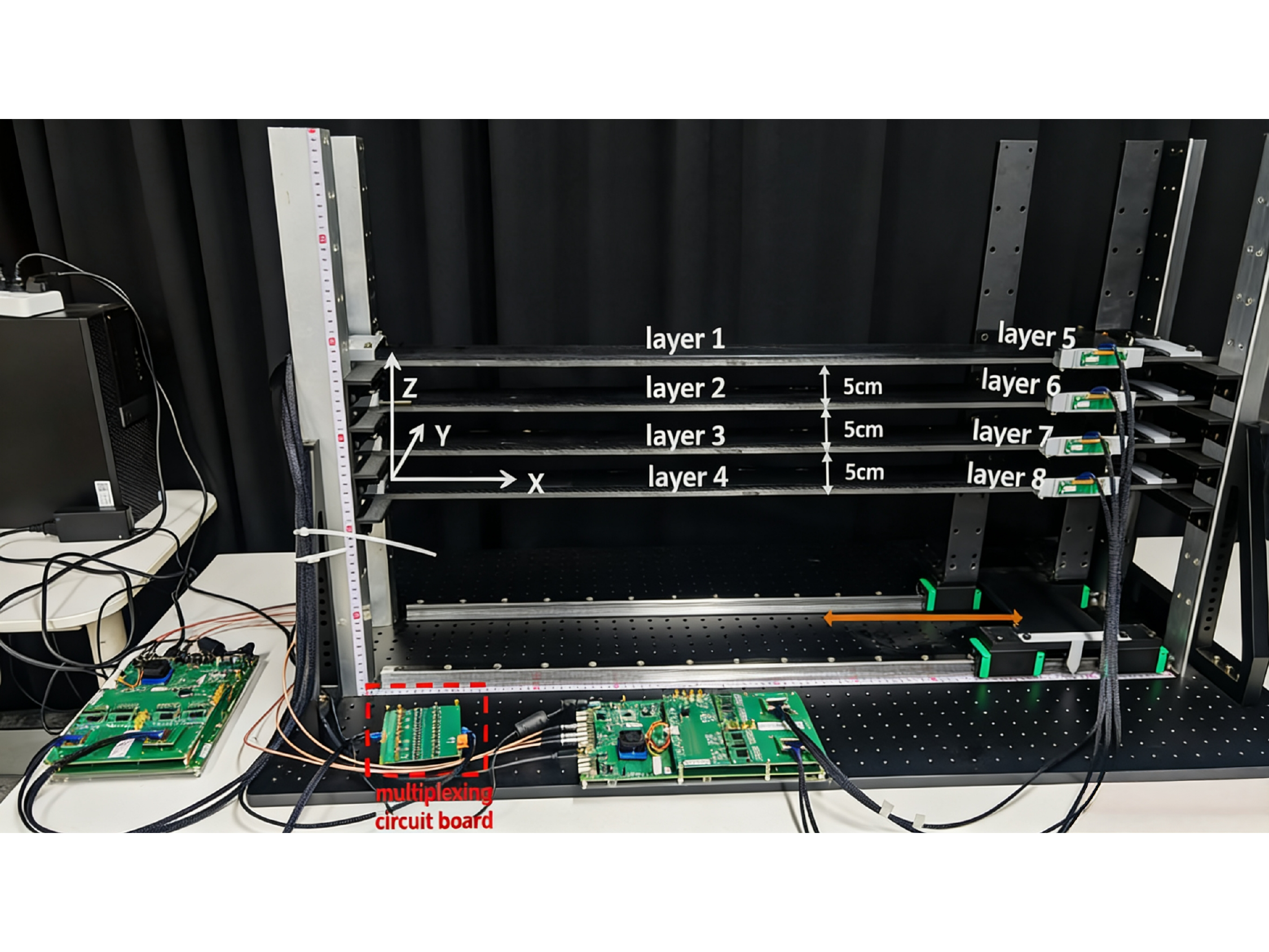}}
\caption{Experimental setup of the SciFi detector system, consisting of four 1~m long modules and four 10~cm short modules.}
\label{fig:detSetup}
\end{figure}

The experimental setup for testing the SciFi modules with the multiplexing circuit is shown in Fig.~\ref{fig:detSetup}. It comprises four 1~m long SciFi modules (Layers~1–4) and four 10~cm short modules (Layers~5–8) arranged perpendicularly to the long modules.

The short modules serve as muon trigger detectors. A trigger is generated when any SiPM channel in the short modules records a signal exceeding 10~p.e.. When a trigger is issued, all SiPM channels of long modules are read out regardless of whether they have detected scintillation photons. The short modules can be moved together along the $x$-axis to scan different positions. The multiplexing board was inserted between the SiPM array and the DT5550W readout board only for Layer~3. During the cosmic-ray muon test, all SiPM channels were biased at 57~V, and the slow shaping time of all Citiroc~1A channels was set to 87.5~ns.

The performance of the long SciFi direct readout module  was evaluated using the same experimental setup described in Ref.~\cite{Li:2025}, where the cluster finding, track reconstruction, detector alignment, as well as the evaluation methods for detection efficiency and position resolution, are explained in detail.

\subsection{Output signals}
\begin{figure}[!htbp]

    \centering
 
        \centering
        \begin{overpic}[width=0.8\hsize]{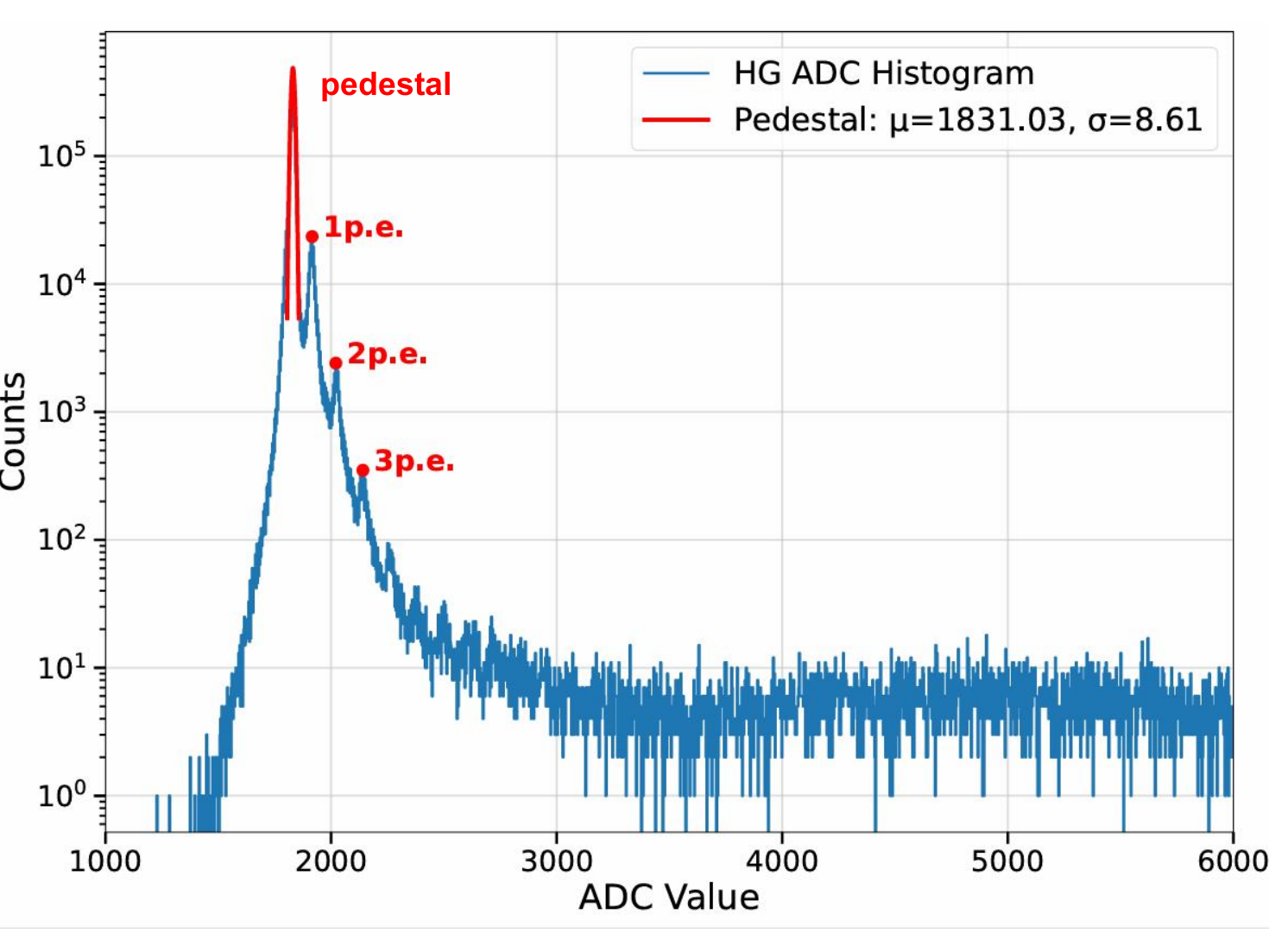}
            \put(10,40){\large\textbf{(a)}}
        \end{overpic}

        \centering
        \begin{overpic}[width=0.8\hsize]{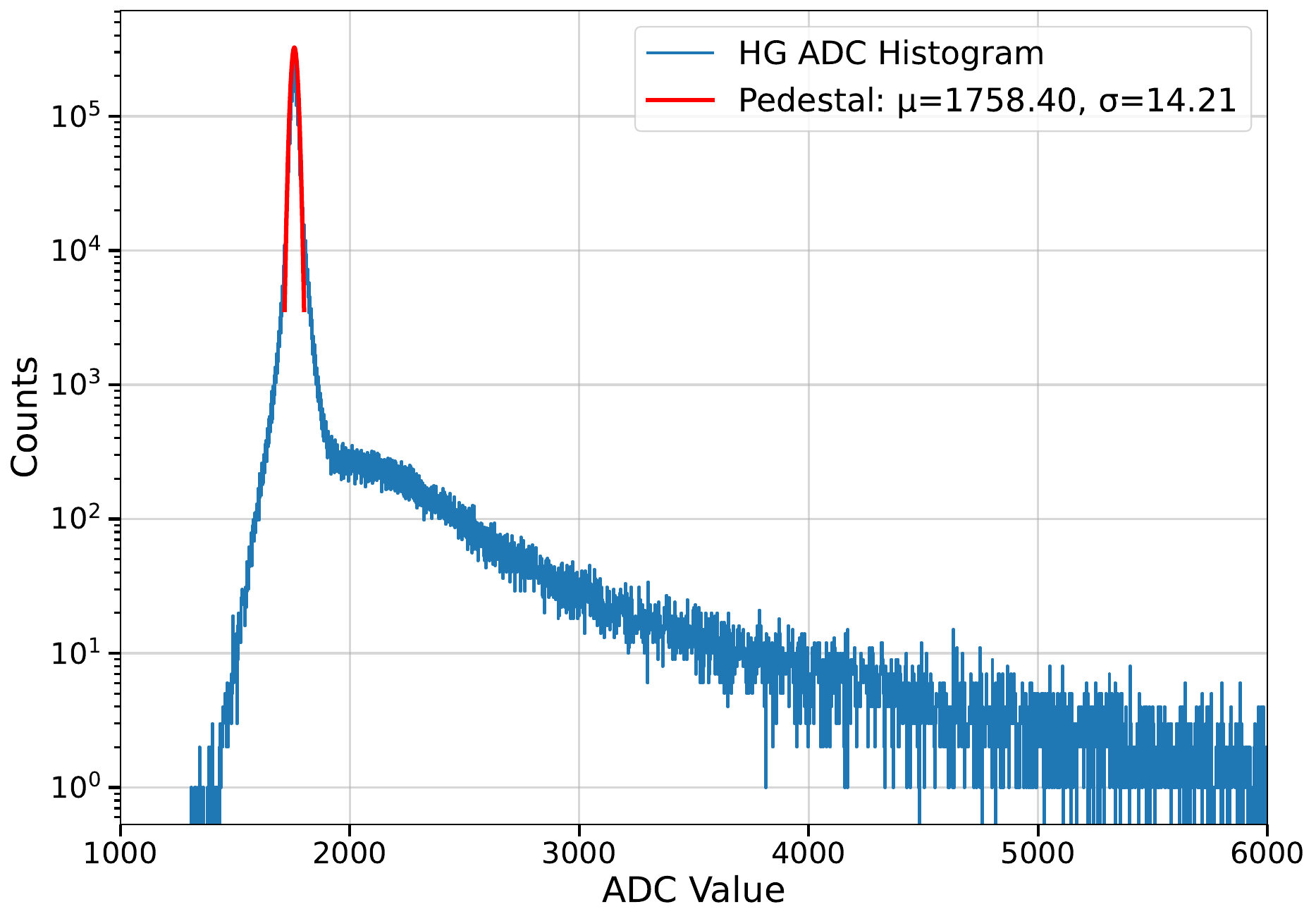}
            \put(10,40){\large\textbf{(b)}}
        \end{overpic}

        \centering
        \begin{overpic}[width=0.9\hsize]{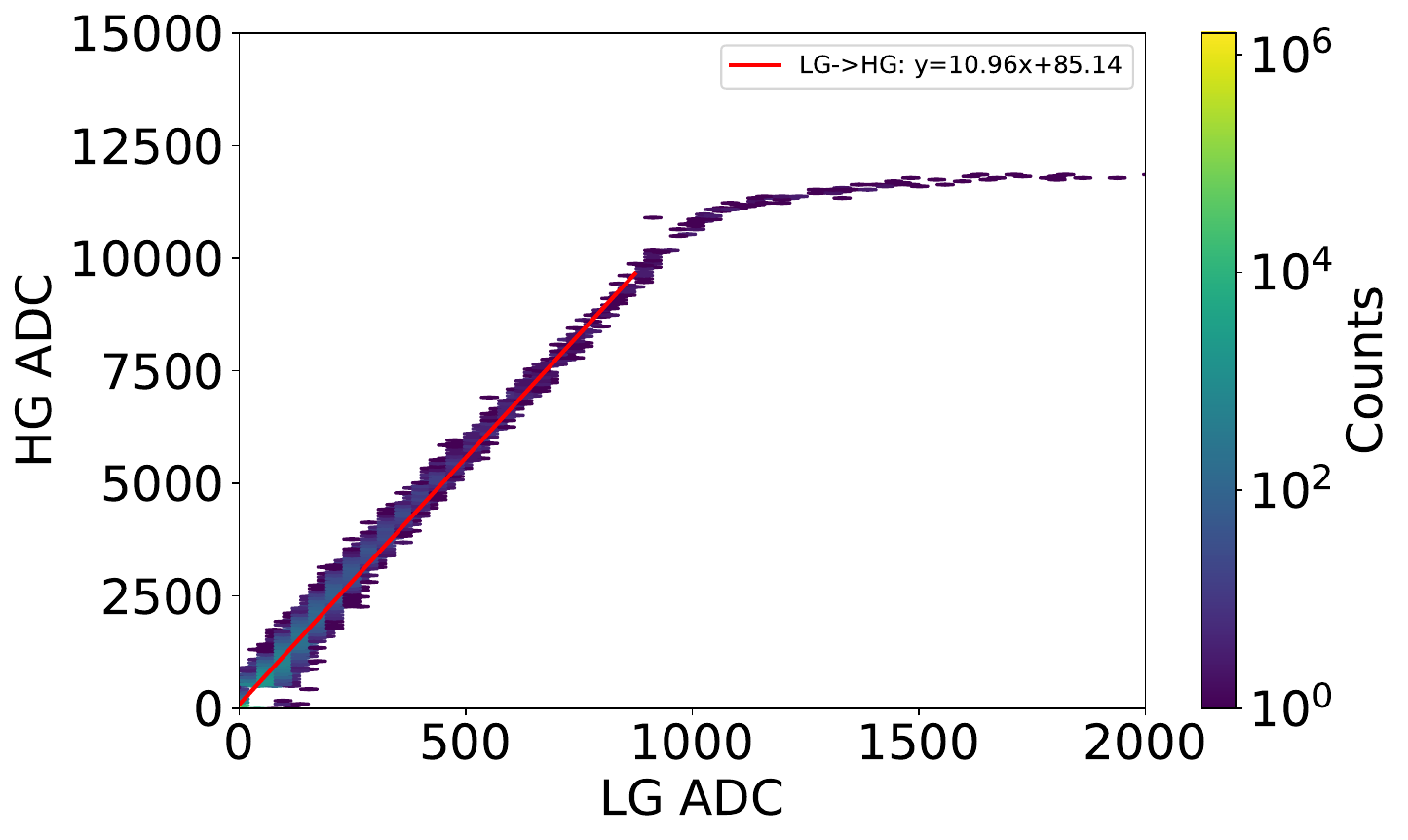}
            \put(20,40){\large\textbf{(c)}}
        \end{overpic}

    \caption{(a) The direct output HG spectrum of one channel in Layer~2, 
    (b) the multiplexed output HG ADC of one channel in Layer~3, 
    and (c) the relationship of LG ADC and HG ADC after pedestal subtraction of one multiplexed channel.}
    \label{fig:output}
\end{figure}

As an example, the HG spectrum of one SiPM channel in Layer~2 is shown in Fig.~\ref{fig:output}(a), where the channel gain can be calibrated as the mean charge difference between adjacent peaks. For a channel connected through the multiplexing circuit, the corresponding HG spectrum is shown in Fig.~\ref{fig:output}(b). In this case, the single-photoelectron peaks are smeared by the multiplexing network, making the standard gain-calibration procedure no longer applicable. After pedestal subtraction, a multiplexed channel is identified as having a valid signal when its HG ADC value exceeds five times the pedestal noise standard deviation, ensuring a reliable discrimination against electronic noise.

For muon events depositing large energy, the HG ADC of a channel may saturate, while its LG response remains linear due to the different gain settings of the Citiroc~1A ASIC. As shown in Fig.~\ref{fig:output}(c), the relationship between the LG and HG ADC values is fitted with a straight line. 
To obtain the signal amplitude of the multiplexed channels in Layer~3, after subtracting the pedestal of HG and LG, the HG ADC is used when it is below 10000, whereas for larger signals, the LG ADC value is used and corrected using the fitted LG–HG linear relation.
All multiplexed channels with valid signals are subsequently decoded to reconstruct the SiPM signals in Layer~3 without applying individual gain calibration.

For future applications, the multiplexed detector can be calibrated by preselecting diodes with similar current–voltage characteristics, tuning the bias voltage of each SiPM to equalize their gains, and performing a charge-injection calibration for all Citiroc~1A channels. These procedures help minimize variations in the photon-detection response across SiPMs and readout channels, thereby improving the uniformity of the multiplexed detector.

\subsection{Detection efficiency and decoding rate}

The successful decoding rate of the multiplexing method is evaluated by comparing the detection efficiencies of the SciFi module between direct and multiplexed readout. To measure the detection efficiency of Layer~3, all short modules except for the one paired with Layer~3 were used to constrain the muon impact position, while the remaining long modules were employed to reconstruct reference tracks for efficiency determination. The tracks were reconstructed with a straight-line fit using the least-squares method after detector alignment, which was performed with the Gauss–Newton algorithm~\cite{Gratton:2007ApproxGN}.

We define $N_{\rm{evt}}$ as the number of muon tracks expected to pass through the fiducial volume of Layer~3 and $N_{\rm{det}}$ as the number of those tracks that produce a valid cluster on Layer~3. The detection efficiency is then calculated as:
\begin{equation}
\varepsilon = \frac{N_{\rm{det}}}{N_{\rm{evt}}}.
\end{equation}

\begin{figure}[!htbp]
\centerline{\includegraphics[width=\hsize]{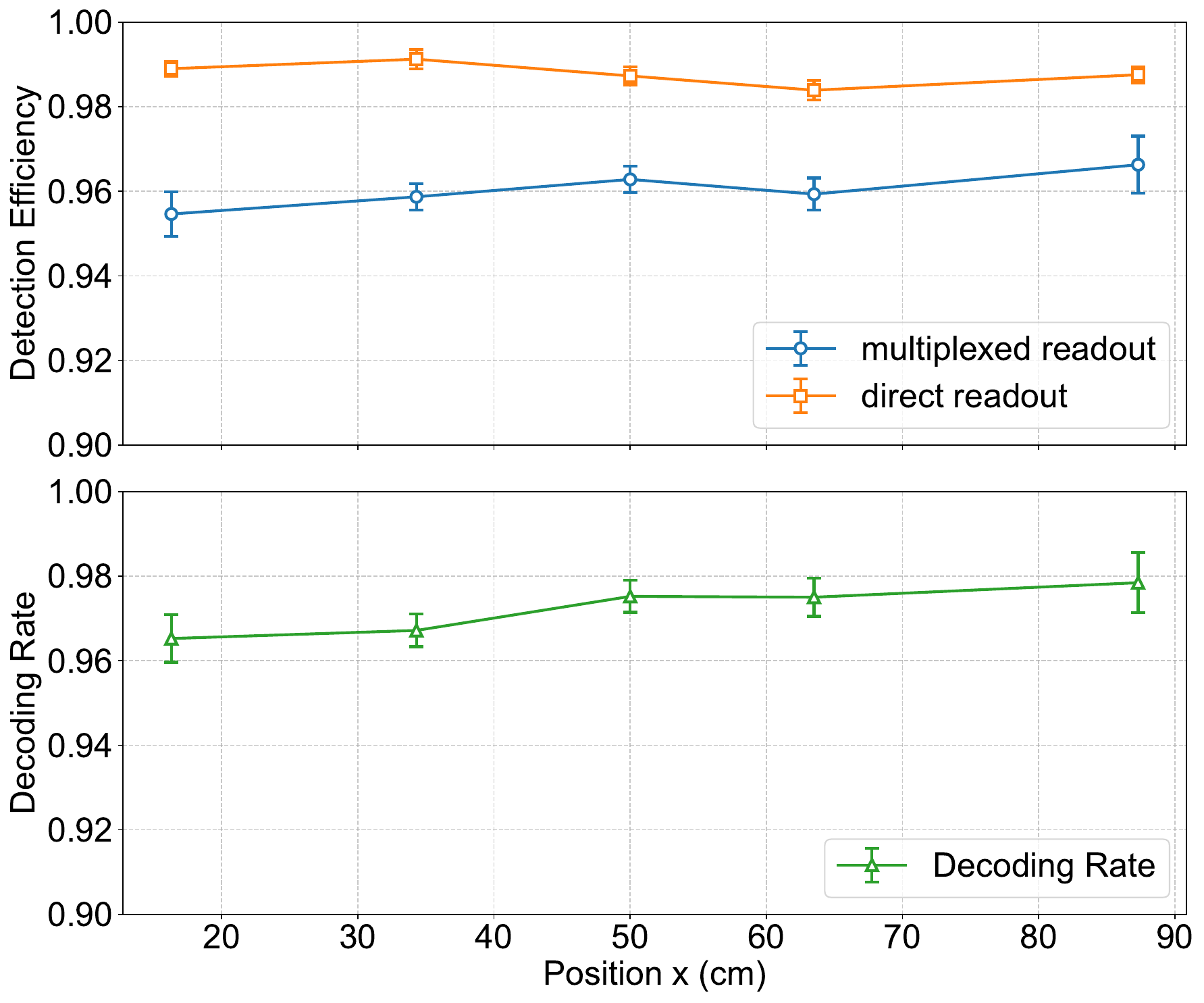}}
\caption{(Top) Comparison of detection efficiency of Layer~3 between multiplexed and direct readout module. (Bottom) The successful decoding ratio of the multiplexing circuits of Layer~3.}
\label{fig:detEff}
\end{figure}

The detection efficiencies between direct and multiplexed  readout were compared at five measurement positions along the $x$-axis, as shown in Fig.~\ref{fig:detEff}. For Layer~3 with direct readout module, the detection efficiency remained above 98\%, while for Layer~3 with multiplexed readout, it stayed above 95\%. The corresponding successful decoding rate was estimated as the ratio of the multiplexed efficiency to the efficiency with direct readout module. As shown in Fig.~\ref{fig:detEff}, the decoding ratio remains above 96\%.

\subsection{Position resolution}

The detector position resolution is a key parameter for developing compact muon tomography systems with good angular resolution. After detector alignment, the position resolution of Layer~3 was evaluated from the residual distribution between the detected cluster on Layer~3 and the extrapolated position of the reconstructed track fitted by the other three long SciFi modules. The residual variance $\sigma_{r}^{2}$ includes contributions from both the intrinsic spatial resolution $\sigma_{d}$ of Layer~3 and the extrapolation uncertainty $\sigma_{f}$, expressed as $\sigma_{r}^{2} = \sigma_{d}^{2} + \sigma_{f}^{2}$. The extrapolation uncertainty is propagated from $\sigma_{d}$ of the layers used in track reconstruction and can be derived as~\cite{Wang:2022TNS}:
\begin{equation}
\begin{aligned}
\sigma_{f,m}^{2} &= \sum_{j\neq m} w_{mj}^{2}\sigma_{d,j}^{2}, \\[3pt]
w_{mj} &= \frac{(z_j-\bar z)(z_m-\bar z)}{\sum_{j\ne m}(z_j-\bar z)^2} + \frac{1}{N-1}, \\[3pt]
\bar z &= \frac{1}{N-1}\sum_{j\ne m} z_j \; ,
\end{aligned}
\end{equation}
where $N$ is the total number of long modules, $z_j$ denotes the vertical position of module $j$ along the $z$-axis, and $w_{mj}$ refers to the weight of each layer uncertainty derived from the error propagation of the straight-line fitting formula. The spatial resolution of each layer was then obtained by solving the set of linear equations for all four modules.

\begin{figure}[!htbp]
\centerline{\includegraphics[width=\hsize]{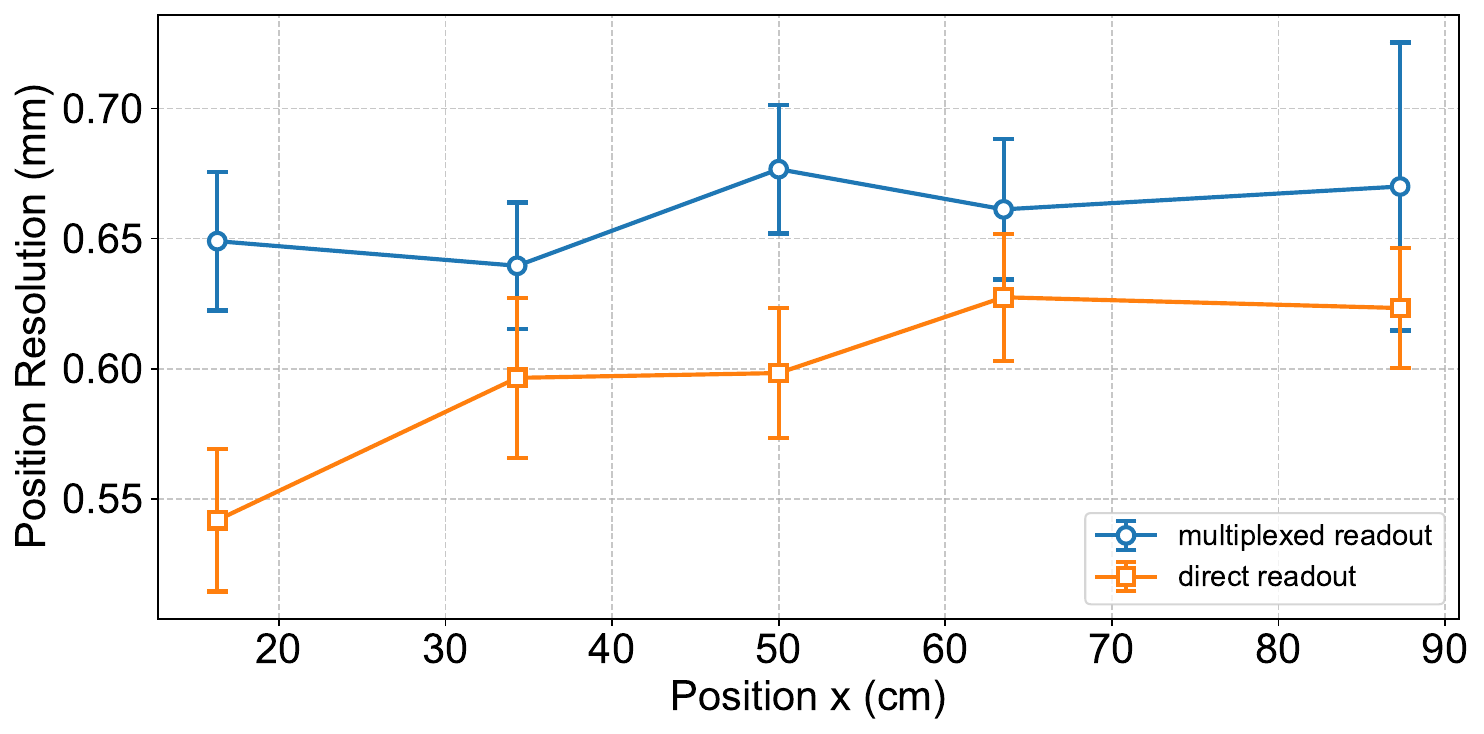}}
\caption{Comparison of position resolution of Layer~3 between multiplexed and direct readout module.}
\label{fig:posRes}
\end{figure}

The position resolutions of Layer~3 with multiplexed and direct readout modules are shown in Fig.~\ref{fig:posRes}. The position resolution of the SciFi module with the multiplexing circuit is approximately 0.65~mm, showing only a slight degradation relative to the direct readout case.

\section{Conclusion}

In this work, a multiplexing method employing diode-based SCD circuits and a position-encoding algorithm was developed and evaluated for the one-dimensional SiPM array readout of a SciFi detector, targeting applications in low event rate muon tomography. The proposed method effectively reduces the number of readout electronic channels while preserving the overall integrity of the detector. Circuit simulations confirmed the feasibility of the multiplexing design and guided the selection of the 1N4007 diode as the preferred component in the circuit implementation.

Joint testing of the multiplexing circuit with a SiPM array demonstrated that the proposed design preserves the SiPM signal shape, exhibits low crosstalk among electronic channels, and maintains a linear response for SiPM signals ranging from $\sim$10 to 122 p.e.. Cosmic-ray muon measurements further showed that the multiplexed SciFi module achieves a detection efficiency above 95\% and a spatial resolution of approximately 0.65~mm, indicating only a slight performance degradation compared with direct readout module.

These results verify that the proposed multiplexing technique offers a cost-effective and scalable readout solution for large-area muon tomography systems, with limited impact on detector performance. The study also demonstrates that the number of electronic channels can be reduced to one third of that required for direct readout, and this reduction factor could be further enhanced in future implementations with SiPM arrays comprising a larger number of channels. For future large-area detectors, the achievable multiplexing factor will be constrained not only by circuit-level design considerations but also by the cumulative baseline noise arising from the simultaneous readout of multiple SiPM channels. Since the noise increases with the number of merged channels, the multiplexing ratio should be optimized to maintain noise fluctuations well below the signal amplitude of the events of interest. Although the multiplexing method was validated using the SciFi detector, it can be readily applied to other scintillator-based detectors employing similar SiPM configurations in muon tomography or related applications.

\bigskip
\textbf{Data Availability Statement} The data that support the findings of this study are openly available in Science Data Bank at https://doi.org/10.57760/sciencedb.33148.


\begin{thebibliography}{99}

\bibitem{Navas:2024}
S. Navas, C. Amsler, T. Gutsche et al., Review of particle physics. Phys. Rev. D 110, 030001 (2024). \href{https://doi.org/10.1103/PhysRevD.110.030001}{doi:10.1103/PhysRevD.110.030001}

\bibitem{Borozdin2003}
K. N. Borozdin, G. E. Hogan, C. Morris et al., Radiographic imaging with cosmic-ray muons. Nature 422, 277 (2003). \href{https://doi.org/10.1038/422277a}{doi:10.1038/422277a}


\bibitem{Bonomi:2020}
G. Bonomi, P. Checchia, M. D’Errico et al., Applications of cosmic-ray muons. Prog. Part. Nucl. Phys. 112, 103768 (2020). \href{https://doi.org/10.1016/j.ppnp.2020.103768}{doi:10.1016/j.ppnp.2020.103768}

\bibitem{Tang2020:MuonSourceReview}
J. Y. Tang, L. P. Zhou, Y. Hong et al., Multidisciplinary research and applications of muon sources. Physics (in Chinese), {\bf 49}: 1000‑1013 (2020). \href{https://doi.org/10.7693/wl20201001}{doi:10.7693/wl20201001}

\bibitem{Tanaka:2023}
H. K. M. Tanaka, C. Bozza, A. Bross et al., Muography. Nat. Rev. Methods Primers 3, 88 (2023). \href{https://doi.org/10.1038/s43586-023-00270-7}{doi:10.1038/s43586-023-00270-7}

\bibitem{Jonkmans:2013NuclearWasteMuon}
G. Jonkmans, V.N.P. Anghel, C. Jewett et al., Nuclear waste imaging and spent fuel verification by muon tomography. Ann. Nucl. Energy 53, 267–273 (2013). \href{https://doi.org/10.1016/j.anucene.2012.09.011}{doi:10.1016/j.anucene.2012.09.011}

\bibitem{Mahon:2018}
D. Mahon, A. Clarkson, S. Gardner et al., First-of-a-kind muography for nuclear waste characterization. Philos. Trans. R. Soc. A 377, 20180048 (2018). \href{https://doi.org/10.1098/rsta.2018.0048}{doi:10.1098/rsta.2018.0048}

\bibitem{TUMUTY:2019}
X. Y. Pan, Y. F. Zheng, Z. Zeng et al., Experimental validation of material discrimination ability of muon scattering tomography at the TUMUTY facility. Nucl. Sci. Tech. 30, 8 (2019). \href{https://doi.org/10.1007/s41365-019-0649-4}{doi:10.1007/s41365-019-0649-4}

\bibitem{Xiao:2018}
S. Xiao, W. B. He, M. C. Lan et al., A modified multi-group model of angular and momentum distribution of cosmic ray muons for thickness measurement and material discrimination of slabs. Nucl. Sci. Tech. 29, 28 (2018). \href{https://doi.org/10.1007/s41365-018-0363-7}{doi:10.1007/s41365-018-0363-7}


\bibitem{Luo:2022}
S. Y. Luo, Y. H. Huang, X. T. Ji et al., Hybrid model for muon tomography and quantitative analysis of image quality. Nucl. Sci. Tech. 33, 81 (2022). \href{https://doi.org/10.1007/s41365-022-01070-6}{doi:10.1007/s41365-022-01070-6}


\bibitem{He:2022MuonANN}
W. He, D. Chang, R. Shi et al., Material discrimination using cosmic ray muon scattering tomography with an artificial neural network. Radiat. Detect. Technol. Methods 6, 254–261 (2022). \href{https://doi.org/10.1007/s41605-022-00319-3}{doi:10.1007/s41605-022-00319-3}

\bibitem{Barnes:2023}
S. Barnes, A. Georgadze, A. Giammanco et al., Cosmic-ray tomography for border security. Instruments 7, 13 (2023). \href{https://doi.org/10.3390/instruments7010013}{doi:10.3390/instruments7010013}

\bibitem{Morishima:2017ghw}
K. Morishima, M. Kuno, A. Nishio et al., Discovery of a big void in Khufu's Pyramid by observation of cosmic-ray muons. Nature 552, 386–390 (2017). \href{https://doi.org/10.1038/nature24647}{doi:10.1038/nature24647}


\bibitem{Liu:2023XianWall}
G. Liu, X. Luo, H. Tian et al., High-precision muography in archaeogeophysics: A case study on Xi'an defensive walls. J. Appl. Phys. 133, 014901 (2023). \href{https://doi.org/10.1063/5.0123337}{doi:10.1063/5.0123337}



\bibitem{Han:2020TunnelMuonFlux}
R. Han, Q. Yu, Z. Li et al., Cosmic muon flux measurement and tunnel overburden structure imaging. J. Instrum. 15, P06019 (2020). \href{https://doi.org/10.1088/1748-0221/15/06/P06019}{doi:10.1088/1748-0221/15/06/P06019}

\bibitem{Cheng:2022}
Y. P. Cheng, R. Han, Z. W. Li, et al., Imaging internal density structure of the Laoheishan volcanic cone with cosmic ray muon radiography. Nucl. Sci. Tech. 33, 88 (2022). \href{https://doi.org/10.1007/s41365-022-01072-4}{doi:10.1007/s41365-022-01072-4}
	
\bibitem{Liu:2024}
G. Liu, K. Yao, F. Niu et al., Deep investigation of muography in discovering geological structures in mineral exploration: A case study of Zaozigou gold mine. Geophys. J. Int. 237, 588–603 (2024). \href{https://doi.org/10.1093/gji/ggae057}{doi:10.1093/gji/ggae057}

\bibitem{MuGridV2:2025}
Y. Tao, Y. Ning, Y. Yuan et al., MuGrid-v2: A novel scintillator detector for multidisciplinary applications. J. Appl. Phys. 138, 024501 (2025). \href{https://doi.org/10.1063/5.0273373}{doi:0.1063/5.0273373}


\bibitem{Lesparre:2012}
N. Lesparre, J. Marteau, Y. Déclais et al., Design and operation of a field telescope for cosmic ray geophysical tomography. Geosci. Instrum. Method. Data Syst. 1, 33–42 (2012). \href{https://doi.org/10.5194/gi-1-33-2012}{doi:10.5194/gi-1-33-2012}


\bibitem{Dong:2018}
J. N. Dong, Y. L. Zhang, Z. Y. Zhang et al., Position-sensitive plastic scintillator detector with WLS-fiber readout. Nucl. Sci. Tech. 29, 117 (2018). \href{https://doi.org/10.1007/s41365-018-0449-2}{doi:10.1007/s41365-018-0449-2}


\bibitem{MuGrid:2022}
H. Yang, G. Luo, T. Yu et al., MuGrid: A scintillator detector towards cosmic muon absorption imaging. Nucl. Instrum. Methods Phys. Res. Sect. A 1042, 167402 (2022). \href{https://doi.org/10.1016/j.nima.2022.167402}{doi:10.1016/j.nima.2022.167402}

\bibitem{Anstasio:2013}
A. Anastasio, F. Ambrosino, D. Basta et al., The MU-RAY detector for muon radiography of volcanoes. Nucl. Instrum. Methods Phys. Res. A 732, 423–426 (2013). \href{https://doi.org/10.1016/j.nima.2013.05.159}{doi:10.1016/j.nima.2013.05.159}


\bibitem{Hu:2020PlasticMuonDetector}
T. Q. Hu, Z. Liang, X. Li et al., Development of a plastic scintillation detector for muon scattering imaging. J. Instrum. 15, P11017 (2020). \href{https://doi.org/10.1088/1748-0221/15/11/P11017}{doi:10.1088/1748-0221/15/11/P11017}



\bibitem{Luo:2022CompactMuonImaging}
X. J. Luo, Q. X. Wang, K. M. Qin et al., Development and commissioning of a compact cosmic ray muon imaging prototype. Nucl. Instrum. Methods Phys. Res. Sect. A 1033, 166720 (2022). \href{https://doi.org/10.1016/j.nima.2022.166720}{doi:10.1016/j.nima.2022.166720}

\bibitem{Wang:2024}
Z. Y. Wang, Y. G. Zhang, X. Li et al., Electronics design for a muon imaging system using triangular plastic scintillators with WLS fiber readouts. J. Instrum. 19, P02033 (2024). \href{https://doi.org/10.1088/1748-0221/19/02/P02033}{doi:10.1088/1748-0221/19/02/P02033}


\bibitem{Clarkson:2014}
A. Clarkson,  D. J. Hamilton, M. Hoek et al., The design and performance of a scintillating-fibre tracker for the cosmic-ray muon tomography of legacy nuclear waste containers. Nucl. Instrum. Methods Phys. Res. A 745, 138–149 (2014). \href{https://doi.org/10.1016/j.nima.2014.01.062}{doi:10.1016/j.nima.2014.01.062}

\bibitem{Anbarjafari:2021}
G. Anbarjafari, A. Anier, E. Avots et al., Atmospheric ray tomography for low-Z materials: implementing new methods on a proof-of-concept tomograph. (2021). \href{https://arxiv.org/abs/2102.12542}{doi:arXiv:2102.12542}

\bibitem{Zhai:2022}
J. Zhai, H. Tang, X. Huang et al., A high-position-resolution trajectory detector system for cosmic ray muon tomography: Monte Carlo simulation. Radiat. Detect. Technol. Methods 6, 244–253 (2022). \href{https://doi.org/10.1007/s41605-022-00313-9}{doi:10.1007/s41605-022-00313-9}

\bibitem{Zhai:2024}
J. Zhai, M. Feng, B. Pan et al., Compact cosmic ray muon scattering imaging system based on plastic scintillating fibers. J. Instrum. 19, P12016 (2024). \href{https://doi.org/10.1088/1748-0221/19/12/P12016}{doi:10.1088/1748-0221/19/12/P12016}


\bibitem{Li:2025}
Y. Li, H. Li, H. Liang et al., Performance evaluation of compact plastic scintillating fiber modules for muon tomography applications. (2025). \href{https://arxiv.org/abs/2509.14674}{doi:arXiv:2509.14674}

\bibitem{Sehgal2025OpticalMultiplexing}
R. Sehgal, V. Jha, Optical multiplexing for channel reduction in scintillator-based muon tomography system. J. Appl. Phys. 138, 244501 (2025). \href{https://doi.org/10.1063/5.0273319}{doi:10.1063/5.0273319}

\bibitem{Park:2022}
H. Park, M. Yi, J. S. Lee., Silicon photomultiplier signal readout and multiplexing techniques for positron emission tomography: a review. Biomed. Eng. Lett. 12, 263–283 (2022). \href{https://doi.org/10.1007/s13534-022-00234-y}{doi:10.1007/s13534-022-00234-y}

\bibitem{Siegel:1996}
S. Siegel, R. W. Silverman, Y. Shao, S. R. Cherry., Simple charge division readouts for imaging scintillator arrays using a multi-channel PMT. IEEE Trans. Nucl. Sci. 43, 1634–1641 (1996). \href{https://doi.org/10.1109/23.507202}{doi:10.1109/23.507202}

\bibitem{Won:2016}
J. Y. Won, G. B. Ko, J. S. Lee., Delay grid multiplexing: simple time-based multiplexing and readout method for silicon photomultipliers. Phys. Med. Biol. 61, 7113–7135 (2016). \href{https://doi.org/10.1088/0031-9155/61/19/7113}{doi:10.1088/0031-9155/61/19/7113}


\bibitem{Wonders:2020}
M. A. Wonders, M. Flaska., Characterization of a mixed-sinusoid multiplexing scheme with silicon photomultipliers and an inorganic scintillator. Nucl. Instrum. Methods Phys. Res. A 959, 163403 (2020). \href{https://doi.org/10.1016/j.nima.2020.163403}{doi:10.1016/j.nima.2020.163403}

\bibitem{Citiroc1A}
Weeroc, Citiroc 1A -- Scientific instrumentation SiPM readout chip (2025). \href{https://www.weeroc.com/read\_out\_chips/citiroc-1a/}{https://www.weeroc.com/read\_out\_chips/citiroc-1a/}


\bibitem{Massari:2016}
R. Massari, D. Caputo, S. Ronchi et al., Low-power charge division circuits for wireless applications based on silicon photomultipliers. IEEE Sens. J. 16, 8214–8219 (2016). \href{https://doi.org/10.1109/JSEN.2016.2573638}{doi:10.1109/JSEN.2016.2573638}


\bibitem{Jung:2021}
J. Jung, Y. Choi, K. Park et al., A diode-based symmetric charge division circuit with grounding path to reduce signal crosstalk and improve detector performance. IEEE Trans. Radiat. Plasma Med. Sci. 6, 788–793 (2021). \href{https://doi.org/10.1109/TRPMS.2021.3112184}{doi:10.1109/TRPMS.2021.3112184}


\bibitem{Olcott:2005}
P. D. Olcott, J. A. Talcott, C. S. Levin et al., Compact readout electronics for position sensitive photomultiplier tubes. IEEE Trans. Nucl. Sci. 52, 21–27 (2005). \href{https://doi.org/10.1109/TNS.2004.843134}{doi:10.1109/TNS.2004.843134}

\bibitem{Wang:2016_SiPMmultiplex}
Z. Wang, X. Sun, K. Lou et al., Design, development and evaluation of a resistor-based multiplexing circuit for a 20 × 20 SiPM array. Nucl. Instrum. Methods Phys. Res. Sect. A 816, 40–46 (2016). \href{https://doi.org/10.1016/j.nima.2016.01.081}{doi:10.1016/j.nima.2016.01.081}

\bibitem{TINA-TI}
Texas Instruments, TINA-TI SPICE-Based Analog Simulation Program (2025). \href{https://www.ti.com.cn/tool/cn/TINA-TI}{https://www.ti.com.cn/tool/cn/TINA-TI}

\bibitem{Acerbi:2019SiPMs}
F. Acerbi, S. Gundacker, Understanding and simulating SiPMs. Nucl. Instrum. Methods Phys. Res. Sect. A 926, 16–35 (2019). \href{https://doi.org/10.1016/j.nima.2018.11.118}{doi:10.1016/j.nima.2018.11.118}

\bibitem{Gratton:2007ApproxGN}
S. Gratton, A. S. Lawless, N. K. Nichols et al., Approximate Gauss–Newton methods for nonlinear least squares problems. SIAM J. Optim. 18, 106–132 (2007). \href{https://doi.org/10.1137/050624935}{doi:10.1137/050624935}

\bibitem{Wang:2022TNS}
Y. Wang, Z. Zhang, S. Liu et al., A high spatial resolution muon tomography prototype system based on Micromegas detector. IEEE Trans. Nucl. Sci. 69, 78–85 (2022). \href{https://doi.org/10.1109/TNS.2021.3137415}{doi:10.1109/TNS.2021.3137415}


\end{thebibliography}
\end{document}